\documentclass[11pt]{article}
\usepackage{epsfig} 
\newcommand{\sect}[1]{ \section{#1} \setcounter{equation}{0} }

\newcommand{\Dslash}{D \! \! \! \! /} 
 
\newcommand{\pslash}{p \! \! \! /}

\newcommand{\half}{\mbox{\small{$\frac{1}{2}$}}} 
\newcommand{\Nc}{N_{\!c}} 
\newcommand{\Nf}{N_{\!f}} 
\newcommand{\NF}{N_{\!F}} 
\newcommand{\NA}{N_{\!A}} 
\newcommand{\Nda}{N^d_{\!A}} 
\newcommand{\Noda}{N^o_{\!A}} 
\newcommand{\MSbar}{\overline{\mbox{MS}}} 

\begin{document}

\title{Three loop $\MSbar$ renormalization of QCD in the maximal abelian gauge}
\author{J.A. Gracey, \\ Theoretical Physics Division, \\ 
Department of Mathematical Sciences, \\ University of Liverpool, \\ P.O. Box
147, \\ Liverpool, \\ L69 3BX, \\ United Kingdom.} 
\date{} 
\maketitle 
\vspace{5cm} 
\noindent 
{\bf Abstract.} We determine the three loop anomalous dimensions of the quark, 
centre and off-diagonal gluons, centre and off-diagonal ghosts and the gauge 
fixing parameters in the maximal abelian gauge for an arbitrary colour group in
the $\MSbar$ renormalization scheme at three loops. We show that the three 
loop $\MSbar$ $\beta$-function emerges from the renormalization of the centre
gluon and also deduce the anomalous dimension of the BRST invariant dimension
two mass operator. Moreover, we demonstrate that in the limit that the 
dimension of the centre of the group tends to zero, the anomalous dimensions of
the quarks, off-diagonal gluons and off-diagonal ghosts tend to those of the 
quarks, gluons and ghosts of the Curci-Ferrari gauge respectively. 

\vspace{-17.5cm}
\hspace{13.5cm}
{\bf LTH 650} 

\newpage 

\sect{Introduction.} 
The multiloop renormalization of quantum chromodynamics (QCD), the quantum
field theory underlying the strong interactions, has now been successfully
determined at four loops in the $\MSbar$ scheme, \cite{1,2,3,4,5,6,7}. Indeed 
the one loop $\beta$-function, \cite{1}, establishes the important property of 
asymptotic freedom. Further, with the need for more accurate theoretical 
results such as the precise way in which the coupling constant runs, higher 
loop corrections proved necessary. Subsequently, the scheme independent two 
loop result was computed in \cite{2} prior to the three loop calculation of 
\cite{4}. Given the large increase in the number of Feynman diagrams with loop 
order and the parallel problem of devising an algorithm to extract the 
divergence structure of difficult four loop master integrals, it was several 
years before the four loop $\beta$-function appeared, \cite{6}. Indeed given 
the complexity of such a calculation, it was only technically possible with the
intense use of the symbolic manipulation programme {\sc Form}, \cite{8}. Though
the three loop result of \cite{4} also used computer technology and the 
{\sc Mincer} algorithm, \cite{9}. There was an underlying thread to all these 
computations which lay in a judicious choice of gauge in which to perform the 
calculation. Although the $\beta$-function is gauge independent, choosing a 
general covariant gauge, say, to carry out the calculations could have resulted
in a large amount of extra unnecessary computation. This was avoided by 
considering the Feynman gauge where the gluon propagator reduces to one term 
proportional to a scalar field propagator. Only after the original Feynman 
gauge calculations were performed were computations with gluon propagators in 
the full covariant gauge subsequently carried out, \cite{3,5,7}. These were 
necessary for other problems aside from justifying the full gauge parameter 
independence of the $\beta$-function. 

For instance, the anomalous dimensions of the fields as functions of the 
covariant gauge parameter, $\alpha$, were required for a variety of composite 
operator renormalizations such as those central to deep inelastic scattering. 
(See, for example, \cite{10,11}.) Also, it has recently been established that 
there is an interesting relation, \cite{12,13}, in respect of the dimension two
BRST invariant operator which could play the role of a gluon mass. In 
\cite{12,13} it was demonstrated that in the Landau gauge its renormalization 
is not independent, being related to the gluon and ghost anomalous dimensions. 
This was observed by an explicit three loop computation in the $\MSbar$ scheme,
\cite{12}. More recently, the explicit renormalization has been determined at 
four loops through the provision of the Landau gauge gluon and ghost anomalous 
dimensions at that order, \cite{14}. Significantly, similar identities for the 
analogous operator exist in other gauges such as the maximal abelian gauge 
(MAG), \cite{15,16}, and in space-time dimensions other than four, \cite{17}. 
Since these dimension two operators have been the subject of intense analytic 
investigation in various gauges in recent years, see, for instance 
\cite{18,19,20,21,22,23,24,25,26,27,28,29,30,31,32} and references therein, due
to their condensation in a non-trivial vacuum, there is a clear need to 
renormalize QCD in this gauge. In particular the {\em explicit} values of all 
the anomalous dimensions are required as the first step in the extension of the
local composite operator (LCO) method for QCD, \cite{18}, to the MAG in various
colour groups. This would thus open up the possibility of extending the
effective potential calculations in the Landau gauge, \cite{18,30}, to 
situations beyond the few one loop $SU(2)$ MAG studies already considered, 
\cite{19,22,26,32,33}. This is the main aim of this article where we will 
perform the full $\MSbar$ renormalization of QCD in the MAG for an arbitrary 
colour group to determine the explicit values of the anomalous dimensions with
the renormalizability of the gauge having been discussed in 
\cite{32,34,35,36,37,38}. Though given the nature of the MAG construction where
the colour group is split into its centre and off-diagonal sectors, we will 
make several assumptions about the group structure which we have checked are at
least valid in $SU(2)$ and $SU(3)$. It is important to note that the only 
previous explicit renormalization of QCD in the MAG was at one loop and for the
specific group $SU(2)$, \cite{25,32,37,38}. 

In referring to the MAG it is important to note at the outset that we are in
fact considering the more general modified MAG as discussed in \cite{22} for 
$SU(2)$. The reason for this is that the true MAG is defined in a similar
fashion to the Landau gauge. However, by minimizing the square of the gauge
potential over only the off-diagonal sector of the the colour group, as opposed
to the full group in the usual covariant gauge situation, it transpires that 
the renormalization of the subsequent gauge fixed Lagrangian is singular. 
Therefore, analogously to the generalized Landau gauge or covariant gauge, a 
covariant gauge parameter, $\alpha$, is introduced which is not to be confused 
with the parameter of the covariant gauges. With this non-zero $\alpha$ one has
the modified MAG and as we will show, it is the renormalization of $\alpha$
itself which becomes singular as $\alpha$~$\rightarrow$~$0$. However, all the
remaining renormalization group functions are finite as 
$\alpha$~$\rightarrow$~$0$ whence one obtains the true MAG anomalous 
dimensions. Moreover, as has been observed before, \cite{19}, the structure of 
the MAG renormalization has connections not only with the Landau gauge but also
with the related non-linear covariant gauge known as the Curci-Ferrari gauge 
introduced in \cite{39}. It will turn out that such connections will also prove
useful for justifying our final three loop $\MSbar$ anomalous dimensions. 

Another motivation for considering the MAG rests in one of the original reasons
why it was introduced. One possibility for the mechanism of confinement is
the condensation of abelian monopoles which clearly originate from the centre
of the colour group, \cite{40,41,42}. In any calculations which seek to focus 
on this supposition, it makes sense to consider a gauge where the centre and 
off-diagonal fields are separately identified in the gauge fixing. Therefore,
by establishing the renormalization structure at three loops in this gauge, one
would expect the results will be useful, say, in any continuum matching one 
might have to do in lattice computations. On a final note we draw attention to 
another gauge in which QCD is renormalized and that is the background field 
gauge where the gauge field is split into a classical and quantum part,
\cite{43,44,45,46}. The latter is regarded as the totally internal quantum 
fluctuation. In addition to the other three loop results referred to earlier, 
QCD has also been renormalized to the same order in this gauge, 
\cite{45,46,47}. The main advantage of the background field gauge is the fact 
that the $\beta$-function emerges from the renormalization of the gluon field. 
In other words one needs only to consider a $2$-point function rather than a 
$3$-point function which considerably simplifies any explicit computation. 
Interestingly, the MAG, where the gluon field is split, but with respect to the
colour property, has an analogous simplification which is that the centre gluon
anomalous dimension is also equivalent to the $\beta$-function, \cite{32}. This
feature will be exploited here to reduce the number of Feynman diagrams we have
to consider to perform the full three loop renormalization.

The paper is organized as follows. In section $2$ we review how the MAG
Lagrangian itself is constructed prior to summarizing the group theory results
which were required for the three loop renormalization. This is a non-trivial
exercise since the colour indices have to be identified either as originating
in the centre of the Lie group or in the off-diagonal sector. The details of 
the full three loop renormalization are discussed in section $3$ where the 
structure of the actual renormalization established with the algebraic 
renormalization formalism, \cite{32}, is reviewed. This section also contains 
the main results of the computation which is the determination of the explicit
values of all renormalization group functions for the MAG. Finally, section $4$
contains concluding remarks and the appendix contains the non-trivial Feynman 
rules used in the calculation.  
 
\sect{Maximal abelian gauge.}
We begin by recalling the essential features of the maximal abelian gauge
fixing which depends on the parameter $\alpha$. First, we note that the colour
group generators are $T^A$ where $1$~$\leq$~$A$~$\leq$~$\NA$ and $\NA$ is the
dimension of the adjoint representation. Thus the group valued gauge field
${\cal A}_\mu$ can be decomposed as  
\begin{equation}
{\cal A}_\mu ~=~ A^A_\mu T^A ~. 
\end{equation} 
In considering the MAG the group generators are split into two sets. Those
corresponding to the generators of the centre of the group, which themselves
form a group, and the remaining set. For notational purposes we will use the
indices $i$, $j$, $k$ and $l$ to denote centre elements and $a$, $b$, $c$ and
$d$ to denote off-diagonal elements. Thus ${\cal A}_\mu$ can alternatively be
decomposed as  
\begin{equation}
{\cal A}_\mu ~=~ A^a_\mu T^a ~+~ A^i_\mu T^i 
\end{equation} 
where we introduce the dimension of the centre by noting that 
$1$~$\leq$~$i$~$\leq$~$\Nda$ and allowing the off-diagonal indices to range
over $1$~$\leq$~$a$~$\leq$~$\Noda$. Clearly 
\begin{equation}
\Nda ~+~ \Noda ~=~ \NA
\end{equation}
and, for instance, in the unitary groups $SU(\Nc)$ we have $\Nda$ $=$ $\Nc-1$
and $\Noda$ $=$ $\Nc(\Nc-1)$. With this notation the QCD Lagrangian in general
is, with the gauge fixing part $L_{\mbox{\footnotesize{gf}}}$ to be specified, 
\begin{equation}
L ~=~ -~ \frac{1}{4} G^A_{\mu\nu} G^{A \, \mu\nu} ~+~ i \bar{\psi} \Dslash 
\psi ~+~ L_{\mbox{\footnotesize{gf}}} 
\end{equation} 
where $G^A_{\mu\nu}$ $=$ $\partial_\mu A^A_\nu$ $-$ $\partial_\nu A^A_\mu$ $+$ 
$g f^{ABC} A^B_\mu A^C_\nu$, $D_\mu$ is the covariant derivative, there are
$\Nf$ flavours of quarks, $\NF$ is the dimension of the fundamental 
representation and $g$ is the coupling constant. For the MAG the indices $A$
are split into the two sectors giving  
\begin{equation}
L ~=~ -~ \frac{1}{4} G^a_{\mu\nu} G^{a \, \mu\nu} ~-~ 
\frac{1}{4} G^i_{\mu\nu} G^{i \, \mu\nu} ~+~ i \bar{\psi} \Dslash \psi ~+~ 
L_{\mbox{\footnotesize{gf}}} 
\label{maglag1} 
\end{equation} 
where now $L_{\mbox{\footnotesize{gf}}}$ is interpreted as the MAG gauge fixing
term. This is constructed, see, for example, \cite{22,32}, in the standard way
by the BRST variation of a specific operator. In the usual covariant gauge 
fixing one uses 
\begin{equation}
L_{\mbox{\footnotesize{gf}}}^{\mbox{\footnotesize{cov}}} ~=~ \delta 
\bar{\delta} \left[ \half A_\mu^A A^{A \, \mu} ~+~ \half \alpha \bar{c}^A c^A 
\right] 
\label{maggf1} 
\end{equation} 
where $\delta$ and $\bar{\delta}$ are the BRST and anti-BRST variations
respectively, $c^A$ is the ghost field and $\bar{c}^A$ is the anti-ghost 
field. In the MAG the gauge fixing term is chosen in a similar way. The
off-diagonal sector is chosen as in the covariant gauge case but the diagonal  
sector is restricted to being in the Landau gauge to fully fix the gauge
overall. It is not instructive to repeat all the additional technical details 
of the gauge fixing which have been discussed previously in \cite{22,32}. 
Therefore, for the MAG we take, \cite{22,32},  
\begin{equation}
L_{\mbox{\footnotesize{gf}}} ~=~ \delta \bar{\delta} \left[ \half
A_\mu^a A^{a \, \mu} ~+~ \half \alpha \bar{c}^a c^a ~+~ \half \zeta 
A_\mu^i A^{i \, \mu} \right] ~+~ ( 1 - \zeta ) \delta \left[ \bar{c}^i 
\partial^\mu A_\mu^i \right] 
\end{equation} 
where the last term is included to ensure one can interpolate the results 
between the MAG and the Landau gauge according to how one chooses the 
additional parameter $\zeta$. For instance, the Landau gauge corresponds to
$\alpha$ $=$ $0$ and $\zeta$ $=$ $1$ and the (modified) MAG is $\alpha$ $\neq$
$0$ but $\zeta$ $=$ $0$. This particular gauge fixing was introduced in 
\cite{22} and we have chosen to work with this version for various reasons. 
First, this Lagrangian has been examined from the algebraic renormalization 
point of view and the Slavnov-Taylor identities have been established. Second, 
and more crucially for the current article, in a computation of the magnitude 
of the three loop MAG renormalization it is important to recognise that
calculating with an arbitrary $\alpha$ and $\zeta$ allows us to check the 
correctness of, say, our programming and resultant renormalization constants. 
In particular the Landau gauge three loop anomalous dimensions ought to 
correctly emerge from the computation prior to specifying the MAG values of the
parameter $\zeta$. This is actually a non-trivial point since we have to 
perform the group theory manipulations for the split group and not the full 
group as one would do in an ordinary covariant gauge fixed Lagrangian where the
Casimir structure resulting from group identities is already well established. 

With the MAG gauge fixing, (\ref{maggf1}), it is elementary to perform the BRST
and anti-BRST variations, which are given by 
\begin{eqnarray} 
\delta A^a_\mu &=& -~ \left( \partial_\mu c^a + g f^{ajc} A^j_\mu c^c
+ g f^{abc} A^b_\mu c^c + g f^{abk} A^b_\mu c^k \right) \nonumber \\ 
\delta c^a &=& g f^{abk} c^b c^k + \frac{1}{2} f^{abc} c^b c^c ~~~,~~~ 
\delta \bar{c}^a ~=~ b^a ~~~,~~~ \delta b^a = 0 ~, \nonumber \\
\delta A^i_\mu &=& -~ \left( \partial_\mu c^i + g f^{ibc} A^b_\mu c^c
\right) ~~~,~~~ \delta c^i ~=~ \frac{1}{2} g f^{ibc} c^b c^c ~~~,~~~ 
\delta \bar{c}^i ~=~ b^i ~~~,~~~ \delta b^i ~=~ 0 
\end{eqnarray} 
and 
\begin{eqnarray} 
\bar{\delta} A^a_\mu &=& -~ \left( \partial_\mu c^a + g f^{ajc} A^j_\mu c^c
+ g f^{abc} A^b_\mu c^c + g f^{abk} A^b_\mu c^k \right) \nonumber \\ 
\bar{\delta} c^a &=& -~ b^a + g f^{abc} c^b \bar{c}^c + g f^{abk} c^b \bar{c}^k
+ g f^{abk} \bar{c}^b c^k \nonumber \\
\bar{\delta} \bar{c}^a &=& g f^{abk} \bar{c}^b \bar{c}^k + \frac{1}{2} g f^{abc}
\bar{c}^b \bar{c}^c ~~~,~~~
\bar{\delta} b^a = -~ g f^{abc} b^b \bar{c}^c - g f^{abk} b^b \bar{c}^k
+ g f^{abk} \bar{c}^b b^k ~, \nonumber \\
\bar{\delta} A^i_\mu &=& -~ \left( \partial_\mu \bar{c}^i + g f^{ibc} 
A^b_\mu \bar{c}^c \right) ~~~,~~~ 
\bar{\delta} c^i ~=~ -~ b^i + g f^{ibc} c^b \bar{c}^c ~~~,~~~ 
\bar{\delta} \bar{c}^i ~=~ \frac{1}{2} g f^{ibc} \bar{c}^b \bar{c}^c ~, 
\nonumber \\
\bar{\delta} b^i &=& -~ g f^{ibc} b^b \bar{c}^c 
\end{eqnarray} 
respectively, to obtain the gauge fixed MAG Lagrangian 
\begin{eqnarray}
L_{\mbox{\footnotesize{gf}}} &=& \frac{\alpha}{2} b^a b^a + b^a \partial^\mu
A^a_\mu + \frac{\bar{\alpha}}{2} b^i b^i + b^i \partial^\mu A^i_\mu 
+ \bar{c}^a \partial^\mu \partial_\mu c^a + \bar{c}^i \partial^\mu 
\partial_\mu c^i \nonumber \\
&& -~ g \, b^a \left[ ( 1 - \zeta ) f^{abk} A^b_\mu A^{k \, \mu}  
- \frac{1}{2} \alpha f^{abc} \bar{c}^b c^c - \alpha f^{abk} \bar{c}^b c^k 
\right] \nonumber \\
&& +~ g \left[ ( 1 - \zeta ) f^{abk} A^a_\mu \bar{c}^k \partial^\mu c^b 
- \zeta f^{abk} A^a_\mu \partial^\mu c^b \bar{c}^k  
- f^{abc} A^a_\mu \partial^\mu c^b \bar{c}^c  
- \zeta f^{abk} A^a_\mu c^b \partial^\mu \bar{c}^k  
\right. \nonumber \\
&& \left. ~~~~~~~ 
-~ f^{abk} \partial^\mu A^a_\mu c^b \bar{c}^k 
- f^{abc} \partial^\mu A^a_\mu \bar{c}^b c^c 
- f^{abk} \partial^\mu A^a_\mu \bar{c}^b c^k 
- ( 2 - \zeta ) f^{abk} A^k_\mu \bar{c}^a \partial^\mu \bar{c}^b \right.
\nonumber \\
&& \left. ~~~~~~~ 
-~ f^{abk} \partial^\mu A^k_\mu \bar{c}^a c^b \right] 
\nonumber \\
&& +~ g^2 \left[ ( 1 - \zeta ) f_d^{acbd} A^a_\mu A^{b \, \mu} \bar{c}^c c^d  
+ ( 1 - \zeta ) f_o^{adcj} A^a_\mu A^{j \, \mu} \bar{c}^c c^d  
+ ( 1 - \zeta ) f_o^{alcj} A^a_\mu A^{j \, \mu} \bar{c}^c c^l 
\right. \nonumber \\
&& \left. ~~~~~~~~ 
+~ ( 1 - \zeta ) f_o^{cjdi} A^i_\mu A^{j \, \mu} \bar{c}^c c^d  
- \frac{\alpha}{4} f_d^{abcd} \bar{c}^a \bar{c}^b c^c c^d  
- \frac{\alpha}{8} f_o^{abcd} \bar{c}^a \bar{c}^b c^c c^d  
- \frac{\alpha}{4} f_o^{abcl} \bar{c}^a \bar{c}^b c^c c^l \right] \nonumber \\ 
\end{eqnarray}   
where we have introduced the term $\half \bar{\alpha} b^i b^i$ to fix the
residual gauge freedom in the full gauge field, \cite{32}. We have also 
simplified the notation by defining 
\begin{equation}
f_d^{ABCD} ~=~ f^{iAB} f^{iCD} ~~~,~~~ 
f_o^{ABCD} ~=~ f^{eAB} f^{eCD} 
\end{equation} 
for the $4$-point vertices. For perturbative computations the auxiliary fields 
$b^a$ and $b^i$ are eliminated by their equations of motion  
\begin{eqnarray} 
b^a &=& -~ \frac{1}{\alpha} \left[ \partial^\mu A^a_\mu - ( 1 - \zeta ) g 
f^{abk} A^b_\mu A^{k \, \mu} - \frac{1}{2} \alpha g f^{abc} \bar{c}^b c^c 
- \alpha g f^{abk} \bar{c}^b c^k \right] \nonumber \\
b^i &=& -~ \frac{1}{\bar{\alpha}} \partial^\mu A^i_\mu 
\end{eqnarray}  
to obtain the MAG Lagrangian in the form we will renormalize it, \cite{32}, 
\begin{eqnarray}
L_{\mbox{\footnotesize{gf}}} &=& -~ \frac{1}{2\alpha} \left( \partial^\mu 
A^a_\mu \right)^2 - \frac{1}{2\bar{\alpha}} 
\left( \partial^\mu A^i_\mu \right)^2 + \bar{c}^a \partial^\mu \partial_\mu c^a
+ \bar{c}^i \partial^\mu \partial_\mu c^i \nonumber \\
&& +~ g \left[ ( 1 - \zeta ) f^{abk} A^a_\mu \bar{c}^k \partial^\mu c^b 
- \zeta f^{abk} A^a_\mu \partial^\mu c^b \bar{c}^k  
- f^{abc} A^a_\mu \bar{c}^b \partial^\mu c^c  
- \zeta f^{abk} A^a_\mu \bar{c}^b \partial^\mu c^k \right. \nonumber \\
&& \left. ~~~~~~~  
-~ \frac{(1-\zeta)}{\alpha} f^{abk} \partial^\mu A^a_\mu A^b_\nu A^{k \, \nu} 
- f^{abk} \partial^\mu A^a_\mu c^b \bar{c}^k 
- \frac{1}{2} f^{abc} \partial^\mu A^a_\mu \bar{c}^b c^c \right. \nonumber \\
&& \left. ~~~~~~~  
-~ ( 2 - \zeta ) f^{abk} A^k_\mu \bar{c}^a \partial^\mu \bar{c}^b 
- f^{abk} \partial^\mu A^k_\mu \bar{c}^b c^c \right] \nonumber \\  
&& +~ g^2 \left[ ( 1 - \zeta ) f_d^{acbd} A^a_\mu A^{b \, \mu} \bar{c}^c c^d  
- \frac{(1-\zeta)^2}{2\alpha} f_o^{akbl} A^a_\mu A^{b \, \mu} A^k_\nu
A^{l \, \nu} 
+ ( 1 - \zeta ) f_o^{adcj} A^a_\mu A^{j \, \mu} \bar{c}^c c^d \right.
\nonumber \\
&& \left. ~~~~~~~~  
-~ \frac{(1-\zeta)}{2} f_o^{ajcd} A^a_\mu A^{j \, \mu} \bar{c}^c c^d 
+ ( 1 - \zeta ) f_o^{ajcl} A^a_\mu A^{j \, \mu} \bar{c}^c c^l 
+ ( 1 - \zeta ) f_o^{alcj} A^a_\mu A^{j \, \mu} \bar{c}^c c^l \right. 
\nonumber \\
&& \left. ~~~~~~~~  
-~ ( 1 - \zeta ) f_o^{cjdi} A^i_\mu A^{j \, \mu} \bar{c}^c c^d  
- \frac{\alpha}{4} f_d^{abcd} \bar{c}^a \bar{c}^b c^c c^d  
- \frac{\alpha}{8} f_o^{abcd} \bar{c}^a \bar{c}^b c^c c^d  
+ \frac{\alpha}{8} f_o^{acbd} \bar{c}^a \bar{c}^b c^c c^d \right. \nonumber \\
&& \left. ~~~~~~~~  
-~ \frac{\alpha}{4} f_o^{abcl} \bar{c}^a \bar{c}^b c^c c^l  
+ \frac{\alpha}{4} f_o^{acbl} \bar{c}^a \bar{c}^b c^c c^l  
- \frac{\alpha}{4} f_o^{albc} \bar{c}^a \bar{c}^b c^c c^l  
+ \frac{\alpha}{2} f_o^{akbl} \bar{c}^a \bar{c}^b c^k c^l \right] ~. 
\label{maglag} 
\end{eqnarray}   
where it is understood that the parameter $\bar{\alpha}$, which is distinct 
from $\alpha$, is set to zero after our renormalization. Having constructed the 
full MAG Lagrangian as a function of the parameters $\alpha$, $\zeta$ and 
$\bar{\alpha}$ we note that the full set of non-zero Feynman rules generated
from (\ref{maglag}) are given in appendix A.  

Since we will be performing our calculation for an arbitrary colour group but
with the group algebra split into centre and off-diagonal sectors, we close 
this section by discussing the main properties of the Lie algebra which were
required. To construct the necessary lemmas we recall that the Lie algebra and 
basic Casimirs for the full group as well as the Jacobi identity are 
\begin{eqnarray}
\left[ T^A , T^B \right] &=& i f^{ABC} T^C \nonumber \\ 
f^{ACD} f^{BCD} &=& C_A \delta^{AB} ~~,~~ T^A T^A ~=~ C_F I ~~,~~ 
\mbox{Tr} \left( T^A T^B \right) ~=~ T_F \delta^{AB} \nonumber \\
0 &=& f^{ABE} f^{CDE} ~+~ f^{BCE} f^{ADE} ~+~ f^{CAE} f^{BDE} ~.  
\label{gpcas1}
\end{eqnarray} 
From the Lie algebra we have that $f^{ijk}$ $=$ $0$ and $f^{ija}$ $=$ $0$
which enshrines the centre property in the algebraic manipulations. So the
second equation of (\ref{gpcas1}) gives the relations
\begin{eqnarray}
C_A \delta^{ab} &=& f^{acd} f^{bcd} ~+~ 2 f^{acj} f^{bcj} \nonumber \\  
C_A \delta^{ij} &=& f^{icd} f^{jcd} ~. 
\label{gpcas2}
\end{eqnarray} 
To proceed we make the assumption that $f^{acd} f^{bcd}$ is proportional to
$\delta^{ab}$ which is certainly true for $SU(2)$ and we have checked it is
also valid in $SU(3)$. In the group theory discussion which follows, it is
important to bear in mind that for groups where this simplifying feature is
not present then one would have to proceed with $f^{acd} f^{bcd}$ being 
proportional to a symmetric rank two tensor. Taking contractions of 
(\ref{gpcas2}) leads to
\begin{equation}
f^{iab} f^{iab} ~=~ \Nda C_A ~~,~~  
f^{abc} f^{abc} ~=~ \left[ \Noda - 2 \Nda \right] C_A 
\end{equation} 
where $\Nda$ is the dimension of the centre and $\Noda$ is the dimension of the
complement of the centre. Hence,
\begin{equation}
f^{icd} f^{jcd} ~=~ C_A \delta^{ij} ~~,~~
f^{acj} f^{bcj} ~=~ \frac{\Nda}{\Noda} C_A \delta^{ab} ~~,~~ 
f^{acd} f^{bcd} ~=~ \frac{[\Noda - 2 \Nda]}{\Noda} C_A \delta^{ab} ~.  
\end{equation} 
With these elementary results we can use the Jacobi identity to establish
several useful relations which were used extensively throughout the computation
\begin{eqnarray} 
f^{apq} f^{bpr} f^{cqr} &=& \frac{[\Noda - 3 \Nda]}{2\Noda} C_A f^{abc} ~~,~~ 
f^{apq} f^{bpi} f^{cqi} ~=~ \frac{\Nda}{2\Noda} C_A f^{abc} 
\nonumber \\ 
f^{ipq} f^{bpr} f^{cqr} &=& \frac{[\Noda - 2 \Nda]}{2\Noda} C_A f^{ibc} ~~,~~  
f^{ipq} f^{bpj} f^{cqj} ~=~ \frac{\Nda}{\Noda} C_A f^{ibc} ~.  
\end{eqnarray}  
For the group generators, we have 
\begin{equation} 
\mbox{Tr} \left( T^a T^b \right) ~=~ T_F \delta^{ab} ~~,~~ 
\mbox{Tr} \left( T^a T^i \right) ~=~ 0 ~~,~~ 
\mbox{Tr} \left( T^i T^j \right) ~=~ T_F \delta^{ij} 
\end{equation} 
and we make the assumption that $T^i T^i$ is proportional to the unit matrix 
which is certainly true for the groups $SU(N)$. Then, from (\ref{gpcas1}) we 
have
\begin{equation}
T^i T^i ~=~ \frac{T_F}{\NF} \Nda I
\end{equation}
after contracting $\mbox{Tr} \left( T^i T^j \right)$. Hence,
\begin{equation}
T^a T^a ~=~ \left[ C_F ~-~ \frac{T_F}{\NF} \Nda \right] I ~.  
\end{equation} 
It therefore follows from the Lie algebra itself that 
\begin{eqnarray}
T^b T^a T^b &=& \left[ C_F ~-~ \frac{C_A}{2} ~-~ \frac{T_F}{\NF} \Nda ~+~
\frac{C_A \Nda}{2\Noda} \right] T^a ~~,~~  
T^i T^a T^i ~=~ \left[ \frac{T_F}{\NF} \Nda ~-~ \frac{C_A \Nda}{2\Noda} \right]
T^a \nonumber \\ 
T^a T^i T^a &=& \left[ \frac{T_F}{\NF} \Noda ~-~ \frac{C_A}{2} \right] 
T^i ~~,~~ T^j T^i T^j ~=~ \frac{T_F}{\NF} \Nda T^i ~.  
\end{eqnarray} 
As a consistency check on these results adding the first pair together recovers
the usual result
\begin{equation}
T^B T^A T^B ~=~ \left[ C_F ~-~ \frac{C_A}{2} \right] T^A 
\end{equation}
for a free off-diagonal index. Summing the final pair is also consistent with 
this result after use of the relation 
\begin{equation}
C_F \NF ~=~ \left[ \Noda ~+~ \Nda \right] T_F 
\end{equation}  
which follows from taking the trace of $T^A T^A$. Next, given the Lie algebra
it is straightforward to construct the useful lemmas 
\begin{eqnarray}
f^{abc} T^b T^c &=& \frac{i[\Noda-2\Nda]}{2\Noda} C_A T^a ~~~,~~~  
f^{abj} T^b T^j ~=~ \frac{i\Nda}{2\Noda} C_A T^a \nonumber \\  
f^{ibc} T^b T^c &=& \frac{i}{2} C_A T^i ~. 
\end{eqnarray} 
Whilst these results proved to be the workhorse for the full three loop
computation as well, it turned out that at three loops it was quicker to 
include additional lemmas to speed up the group theory computation of our 
{\sc Form} programmes. These were derived from several applications of the 
Jacobi identities and are 
\begin{eqnarray}
f^{apq} f^{brs} f^{qms} f^{cmt} f^{prt} &=& 0 \nonumber \\
f^{apq} f^{bjs} f^{qms} f^{cmt} f^{pjt} &=& 0 \nonumber \\
f^{apq} f^{brs} f^{qjs} f^{cjt} f^{prt} &=& 0 \nonumber \\
f^{apq} f^{brj} f^{qmj} f^{cmk} f^{prk} &=& \frac{{\Nda}^2 C_A^2}{4{\Noda}^2}
f^{abc} \nonumber \\
f^{apj} f^{brs} f^{jms} f^{imt} f^{prt} &=& \frac{\Nda[\Noda-2\Nda]C_A^2}
{4{\Noda}^2} f^{abi} \nonumber \\
f^{apj} f^{bks} f^{jms} f^{imt} f^{pkt} &=& \frac{{\Nda}^2 C_A^2}{{\Noda}^2}
f^{abi} \nonumber \\
f^{apq} f^{brs} f^{qms} f^{imt} f^{prt} &=& 0 \nonumber \\
f^{apj} f^{bks} f^{jms} f^{cmt} f^{pkt} &=& \frac{{\Nda}^2 C_A^2}{4{\Noda}^2}
f^{abc} 
\end{eqnarray} 
where we note that the indices $m$, $p$, $q$, $r$, $s$ and $t$ are also 
regarded as off-diagonal. Given the structure of the three loop Green's 
functions these relations were sufficient for handling the group theory 
associated with the gluon $2$-point functions. In that case for any Feynman 
diagram one has at most six structure functions contracted together with two 
free external group indices. However, in the renormalization of the $A^a_\mu 
\bar{c}^i c^b$ vertex at most seven structure functions are contracted together
with three free group indices. For this case the Green's function was 
multiplied by an additional structure function to leave a scalar group string 
to be simplified. The route to achieving this, aside from applying the rules 
discussed so far, was to follow the approach used for the structure constants 
of the full group in that in that case they correspond to the group generators 
in the adjoint representation. In other words one replaces the structure 
constants by 
\begin{equation}
\left( T^A_{\mbox{\footnotesize{adj}}} \right)_{BC} ~=~ -~ i f^{ABC} 
\end{equation}  
and then applies the usual Lie algebra properties to 
$T^A_{\mbox{\footnotesize{adj}}}$ with the proviso that one evaluates
identities in the adjoint representation. For instance, the result
\begin{equation}
T^B T^C T^A T^B T^C ~=~ \left( C_F - C_A \right) \left( C_F - \half C_A \right) 
T^A 
\end{equation}
implies 
\begin{equation} 
T^B_{\mbox{\footnotesize{adj}}} T^C_{\mbox{\footnotesize{adj}}} 
T^A_{\mbox{\footnotesize{adj}}} T^B_{\mbox{\footnotesize{adj}}} 
T^C_{\mbox{\footnotesize{adj}}} ~=~ 0 ~. 
\end{equation}  
For the MAG calculation one can also use this strategy provided one appreciates
that the structure constants with two or more centre indices are identically 
zero which means that the non-trivial structure constants have at least two
off-diagonal indices. These are therefore regarded as the matrix indices which
leaves the third index as the free generator index and this can either be 
centre or off-diagonal. More significantly one must be careful in regarding 
these objects as nothing more than matrices and not as representations of the 
group generators since the matrix indices are only elements of the off-diagonal
sector which is not closed in the group sense. In light of this to 
differentiate from the adjoint representation of the full group we therefore 
choose to define the analogous object $S^A$ by 
\begin{equation}
\left( S^a \right)_{bc} ~=~ f^{abc} ~~,~~ 
\left( S^i \right)_{bc} ~=~ f^{ibc} 
\end{equation} 
where the two matrix indices will always be off-diagonal. Subsequently, given
all the previous lemmas it only remains to resolve objects of the form
\begin{equation}
\mbox{tr} \left( S^A S^B S^C S^D \right) 
\mbox{tr} \left( S^A S^B S^C S^D \right)
\end{equation}
and
\begin{equation}
\mbox{tr} \left( S^A S^B S^C S^D S^A S^B S^C S^D \right) 
\end{equation}
where the trace is the usual matrix trace and we use $\mbox{tr}$ in
contradistinction to the $\mbox{Tr}$ of the full group. Such structures are 
known to occur in higher loop calculations in QCD when $S^A$ is formally 
replaced by $T^A$, \cite{6}, but not until four loops where they arise only in 
terms involving the simple pole in $\epsilon$ where 
$d$~$=$~$4$~$-$~$2\epsilon$. Therefore, to ensure renormalizability they either
have to vanish or cancel since at three loops they can potentially occur in the
double and triple poles in $\epsilon$ as well as the simple one. As these two 
structures emerge with the summed indices in various combinations of centre and
off-diagonal indices, it is appropriate to relate them to a common term via the
relations
\begin{eqnarray}
\mbox{tr} \left( S^i S^j S^k S^l \right) 
\mbox{tr} \left( S^i S^j S^k S^l \right) &=& 
-~ \mbox{tr} \left( S^i S^j S^k S^d \right) 
\mbox{tr} \left( S^i S^j S^k S^d \right) ~+~ \left[ 6 \Nda + \Noda \right] 
\frac{{\Nda}^3 C_A^4}{4{\Noda}^3} \nonumber \\ 
\mbox{tr} \left( S^i S^j S^k S^d \right) 
\mbox{tr} \left( S^i S^j S^k S^d \right) &=& 
-~ \mbox{tr} \left( S^i S^j S^c S^d \right) 
\mbox{tr} \left( S^i S^j S^c S^d \right) \nonumber \\
&& -~ \left[ 4 {\Nda}^2 - {\Noda}^2 \right] \frac{{\Nda}^2 C_A^4}{8{\Noda}^3} 
\nonumber \\ 
\mbox{tr} \left( S^i S^j S^c S^d \right) 
\mbox{tr} \left( S^i S^j S^c S^d \right) &=& 
-~ \mbox{tr} \left( S^i S^b S^c S^d \right) 
\mbox{tr} \left( S^i S^b S^c S^d \right) \nonumber \\ 
&& +~ \mbox{tr} \left( S^i S^b S^c S^d S^i S^b S^c S^d \right) ~+~ 
\mbox{tr} \left( S^i S^j S^c S^d S^i S^j S^c S^d \right) \nonumber \\
&& -~  \left[ 5 {\Nda}^2 - 4 \Nda \Noda + {\Noda}^2 \right] 
\frac{\Nda C_A^4}{8{\Noda}^3} 
\end{eqnarray}
and 
\begin{equation} 
\mbox{tr} \left( S^i S^j S^c S^d S^i S^j S^c S^d \right) ~=~ 0 
\end{equation} 
which are readily established by use of the Lie algebra and the Jacobi
identity. In the actual renormalization of the $A^a_\mu \bar{c}^i c^b$ vertex 
this leaves the two as yet unevaluated structures as  
$\mbox{tr} \left( S^i S^b S^c S^d \right) 
\mbox{tr} \left( S^i S^b S^c S^d \right)$ and  
$\mbox{tr} \left( S^i S^b S^c S^d S^i S^b S^c S^d \right)$. It turns out that
when the pole parts of all the Feynman diagrams for this renormalization are
added up then the coefficients of these structures is finite. 

\sect{Renormalization.} 
Having derived the MAG Lagrangian we now turn to the details of its
renormalization. First, in renormalizing a renormalizable quantum field theory
one ordinarily introduces renormalization constants for all the fields and
parameters in the Lagrangian. For field theories possessing symmetries such as
a gauge symmetry these renormalization constants are not necessarily all
independent. The underlying symmetry can constain several or more to be 
related. To determine such relations, one can apply techniques such as 
algebraic renormalization, \cite{48}, which ensures the Lagrangian is stable 
under quantum corrections. In \cite{32} this approach has been applied to 
(\ref{maglag}) and several interesting relations emerge. For instance, it turns
out that the anomalous dimension of the centre gluons is proportional to the 
QCD $\beta$-function. This is a useful result since for this gauge fixed 
version of QCD it means that one does not have to renormalize a $3$-point 
vertex to determine the known three loop $\beta$-function of \cite{4}. Instead 
one needs only to consider the centre gluon $2$-point function. From a 
practical computational point of view this is a significant observation which 
we exploit later. Moreover, a similar property is also present in the 
background field gauge where the anomalous dimension of the background gluon is
simply related to the coupling constant renormalization, \cite{45,46}. Although
we are a priori aware of the relation of the centre gluon renormalization to 
that of the coupling constant renormalization in the MAG, in defining our 
renormalization constants we choose at the outset to leave this result to 
emerge in the computation rather than put restrictions on the initial setup. 
Therefore, we define the renormalization constants as 
\begin{eqnarray} 
A^{a \, \mu}_{\mbox{\footnotesize{o}}} &=& \sqrt{Z_A} \, A^{a \, \mu} ~~,~~ 
A^{i \, \mu}_{\mbox{\footnotesize{o}}} ~=~ \sqrt{Z_{A^i}} \, A^{i \, \mu} ~~,~~ 
c^a_{\mbox{\footnotesize{o}}} ~=~ \sqrt{Z_c} \, c^a ~~,~~ 
\bar{c}^a_{\mbox{\footnotesize{o}}} ~=~ \sqrt{Z_c} \, \bar{c}^a ~, \nonumber \\
c^i_{\mbox{\footnotesize{o}}} &=& \sqrt{Z_{c^i}} \, c^i ~~,~~ 
\bar{c}^i_{\mbox{\footnotesize{o}}} ~=~ \frac{\bar{c}^i}{\sqrt{Z_{c^i}}} ~~,~~ 
\psi_{\mbox{\footnotesize{o}}} ~=~ \sqrt{Z_\psi} \psi ~, \nonumber \\  
g_{\mbox{\footnotesize{o}}} &=& \mu^\epsilon Z_g \, g ~~,~~ 
\alpha_{\mbox{\footnotesize{o}}} ~=~ Z^{-1}_\alpha Z_A \, 
\alpha ~~,~~ 
\bar{\alpha}_{\mbox{\footnotesize{o}}} ~=~ Z^{-1}_{\alpha^i} Z_{A^i} \, 
\bar{\alpha} ~~,~~ \zeta_{\mbox{\footnotesize{o}}} ~=~ Z_\zeta \zeta 
\label{rencons} 
\end{eqnarray} 
where $\mu$ is the renormalization scale introduced to ensure the coupling
constant is dimensionless in $d$ dimensions, the subscript 
${}_{\mbox{\footnotesize{o}}}$ denotes the bare quantity and the subscript $i$ 
in the field subscripts of the renormalization constants is included to 
indicate that they correspond to centre objects and there is clearly no 
summation over repeated indices in this instance. In writing down
(\ref{rencons}) from \cite{32} we have chosen, by contrast to $A^i_\mu$, to 
encode the structure of the centre ghost renormalization. In particular the 
anti-centre ghost and centre ghost renormalizations are, contrary to the usual
covariant gauge ghost renormalization, inverses of each other and not equal.
This property emerges from the algebraic renormalization analysis, \cite{32}.
From a practical point of view this means that the centre ghost $2$-point
function cannot be used to determine $Z_{c^i}$ since it would be finite, 
\cite{32}. Instead to find $Z_{c^i}$ one has to renormalize the $3$-point 
$A^a_\mu \bar{c}^i c^b$ vertex once the coupling constant and off-diagonal 
gluon and off-diagonal ghost wave function renormalizations have been 
determined at that particular loop order. Therefore, in the MAG one has still 
at least one $3$-point function renormalization to perform. 

However, the benefit in determining $Z_{c^i}$ rests in the fact that the
dimension two BRST invariant operator   
\begin{equation}
{\cal O} ~=~ \half A^a_\mu A^{a \, \mu} ~+~ \alpha \bar{c}^a c^a 
\end{equation}  
possesses an interesting renormalization structure, \cite{15,32}. It transpires
that its anomalous dimension is not independent but satisfies 
\begin{equation}
\gamma_{\cal O}(a) ~=~ -~ \frac{\beta(a)}{a} ~+~ \gamma_{c^i}(a) 
\label{opid} 
\end{equation} 
where 
\begin{equation}
a ~=~ \frac{g^2}{16\pi^2} 
\end{equation} 
and for completeness its associated renormalization constant is defined as 
\begin{equation}
{\cal O}_{\mbox{\footnotesize{o}}} ~=~ Z_{\cal O} \, {\cal O} ~.  
\end{equation}  
Therefore, it will be straightforward to deduce $\gamma_{\cal O}(a)$ from
explicit knowledge of $\gamma_{c^i}(a)$. This is one of the key results 
required for a two loop extension of the LCO method to the condensation of 
${\cal O}$ in the MAG and will be one of the main results of the article. That 
such a relation is present in the MAG is not specific to this gauge. A similar 
relation exists in the Landau gauge, \cite{12,13}, for the analogous operator 
where the indices range over the full colour group. We have also introduced a 
renormalization constant in (\ref{rencons}) for the interpolating parameter 
$\zeta$. However, since we are only interested in the renormalization of the 
MAG itself which corresponds to the fixed point value of $\zeta$~$=$~$0$ it 
turns out that for the MAG renormalization the explicit form of $Z_\zeta$ is 
not required since it will always be multiplied by zero and there are no 
singularities in $\zeta$ in the Feynman rules. 

{\begin{table}[ht] 
\begin{center} 
\begin{tabular}{|c||c|c|c|c|c|} 
\hline 
Green's function & One loop & Two loop & Three loop & Total \\ 
\hline 
$ A^a_\mu \, A^b_\nu$ & $~6$ & $131$ & $~6590$ & $~6727$ \\ 
$ A^i_\mu \, A^j_\nu$ & $~3$ & $~54$ & $~2527$ & $~2584$ \\ 
$ c^a \, {\bar c}^b$ & $~3$ & $~81$ & $~4006$ & $~4090$ \\ 
$ \psi \, {\bar \psi}$ & $~2$ & $~27$ & $~~979$ & $~1008$ \\ 
$ A^a_\mu \, {\bar c}^i \, c^b$ & $~5$ & $287$ & $22621$ & $22913$ \\ 
\hline 
Total & $19$ & $580$ & $36723$ & $37322$ \\  
\hline 
\end{tabular} 
\end{center} 
\begin{center} 
{Table 1. Number of Feynman diagrams for each Green's function for the MAG
renormalization.} 
\end{center} 
\end{table}}  

We now turn to the technical details of the renormalization of (\ref{maglag})
at three loops in the $\MSbar$ scheme. First, the renormalization group
functions we will determine are deduced from the explicit respective
renormalization constants themselves for the MAG, via 
\begin{eqnarray}
\gamma_A(a) &=& \beta(a) \frac{\partial}{\partial a} \ln Z_A ~+~ \alpha 
\gamma_\alpha(a) \frac{\partial}{\partial \alpha} \ln Z_A \nonumber \\
\gamma_\alpha(a) &=& \left[ \beta(a) \frac{\partial}{\partial a} 
\ln Z_\alpha ~-~ \gamma_A(a) \right] \left[ 1 ~-~ \alpha 
\frac{\partial}{\partial \alpha} \ln Z_\alpha \right]^{-1} \nonumber \\ 
\gamma_{A^i}(a) &=& \beta(a) \frac{\partial}{\partial a} \ln Z_{A^i} ~+~ \alpha 
\gamma_\alpha(a) \frac{\partial}{\partial \alpha} \ln Z_{A^i} \nonumber \\
\gamma_{\alpha^i}(a) &=& \beta(a) \frac{\partial}{\partial a} 
\ln Z_{\alpha^i} ~+~ \alpha \gamma_\alpha(a) \frac{\partial}{\partial \alpha} 
\ln Z_{\alpha^i} - \gamma_{A^i}(a) \nonumber \\
\gamma_c(a) &=& \beta(a) \frac{\partial}{\partial a} \ln Z_c ~+~ \alpha 
\gamma_\alpha(a) \frac{\partial}{\partial \alpha} \ln Z_c \nonumber \\
\gamma_{c^i}(a) &=& \beta(a) \frac{\partial}{\partial a} \ln Z_{c^i} ~+~ \alpha 
\gamma_\alpha(a) \frac{\partial}{\partial \alpha} \ln Z_{c^i} \nonumber \\
\gamma_\psi(a) &=& \beta(a) \frac{\partial}{\partial a} \ln Z_\psi ~+~ \alpha 
\gamma_\alpha(a) \frac{\partial}{\partial \alpha} \ln Z_\psi \nonumber \\
\gamma_{\cal O}(a) &=& \beta(a) \frac{\partial}{\partial a} \ln Z_{\cal O} ~+~ 
\alpha \gamma_\alpha(a) \frac{\partial}{\partial \alpha} \ln Z_{\cal O} 
\label{gamdef} 
\end{eqnarray} 
where, similar to the Curci-Ferrari gauge, we have not assumed that 
$Z_\alpha$~$=$~$1$. Though we have set $\bar{\alpha}$~$=$~$0$ and 
$\zeta$~$=$~$0$. Since these are the renormalization group functions we
require, we will therefore renormalize the centre and off-diagonal gluon
$2$-point function, the off-diagonal ghost $2$-point function, the quark
$2$-point function and the $A^a_\mu \bar{c}^i c^b$ $3$-point function all at
three loops. For these Green's functions the Feynman diagrams were generated
with the {\sc Qgraf} package, \cite{49}, and the specific number of diagrams at
each loop order and Green's function are summarized in Table $1$. By contrast, 
to indicate the magnitude of the MAG renormalization using the Feynman rules of
the appendix, we have provided a similar diagram count for the Curci-Ferrari 
gauge three loop renormalization of \cite{12} in Table $2$. To proceed we 
convert the {\sc Qgraf} output format to the electronic notation used by the 
{\sc Mincer} algorithm, \cite{9}, as written in the symbolic manipulation 
package {\sc Form}, \cite{8,50}, in terms of diagram topology and internal 
momentum routing. The {\sc Mincer} algorithm is then applied to all $37322$ 
Feynman diagrams required for the full renormalization. Though it ought to be 
noted that given that the main group theory is carried out prior to determining
the divergence structure of a diagram the value of a graph could be zero purely
from group considerations. For instance, when one has a one loop gluonic 
self-energy subgraph anywhere where one external leg of the subgraph is in the 
centre of the group and the other in the off-diagonal sector, then that graph 
is trivially zero since $f^{icd} f^{bcd}$~$=$~$0$. To appreciate the benefit of
the centre gluon anomalous dimension relation to the $\beta$-function, if such
a relation did not exist then one would have to compute a $3$-point function in
addition to the ones listed in Table $1$. The easiest one from a computer 
algebra point of view is the quark gluon vertex. For an off-diagonal gluon the 
figures for the corresponding first three columns of Table $1$ are $5$, $217$ 
and $13108$, and $3$, $137$ and $8150$ for a centre gluon quark vertex.  

{\begin{table}[ht] 
\begin{center} 
\begin{tabular}{|c||c|c|c|c|c|} 
\hline 
Green's function & One loop & Two loop & Three loop & Total \\ 
\hline 
$ A^A_\mu \, A^B_\nu$ & $3$ & $19$ & $~282$ & $~304$ \\ 
$ c^A \, {\bar c}^B$ & $1$ & $~\,9$ & $~124$ & $~134$ \\ 
$ \psi \, {\bar \psi}$ & $1$ & $~\,6$ & $~~\,79$ & $~~\,86$ \\ 
$ A^A_\mu {\bar \psi} \, \psi$ & $2$ & $33$ & $~697$ & $~732$ \\ 
\hline 
Total & $7$ & $67$ & $1182$ & $1256$ \\  
\hline 
\end{tabular} 
\end{center} 
\begin{center} 
{Table 2. Number of Feynman diagrams for each Green's function for the 
Curci-Ferrari gauge renormalization.} 
\end{center} 
\end{table}}  

To deduce the renormalization constants themselves for each Green's functions,
we apply the procedure discussed in \cite{5}. Here one computes the Green's 
functions in terms of the bare parameters, $g_{\mbox{\footnotesize{o}}}$, 
$\alpha_{\mbox{\footnotesize{o}}}$, $\bar{\alpha}_{\mbox{\footnotesize{o}}}$ 
and $\zeta_{\mbox{\footnotesize{o}}}$. The renormalized values are introduced 
by the definitions (\ref{rencons}) and iteratively by loop order the 
renormalization constants are fixed by demanding that the overall infinity 
remaining is absorbed by the renormalization constant associated with that 
particular Green's function. As the {\sc Mincer} algorithm is based on 
dimensional regularization in $d$~$=$~$4$~$-$~$2\epsilon$ dimensions, we have 
absorbed all the poles in $\epsilon$ using the modified minimal subtraction 
scheme. 

From a practical computing point of view we organised the one and two loop
renormalization in a different way from the three loop computation. For the
former we retained an arbitrary $\alpha$, $\bar{\alpha}$ and $\zeta$ in the
extraction of the renormalization constants. However, for the three loop case,
due to the increase in the number of actual algebraic terms in a Feynman
diagram to be evaluated, due to the presence of the parameters in the 
propagators and vertices, we chose to fix $\zeta$ to be $0$ or $1$ and 
$\bar{\alpha}$~$=$~$0$ when the Feynman rules were substituted. This speeded up
the computation significantly and avoided very large intermediate {\sc Form}
files which are generated. Running our code in the Landau gauge first allowed
us to check the programme was performing correctly before generating the
explicit value of the Feynman graph in the MAG. Moreover, at three loops one
does not need to be concerned about the renormalization of the bare $\zeta$
in this approach since any corrections to this would only appear at three
loops. 

Having summarized the details of the computation we now record the explicit
results. Rather than present the renormalization constants themselves, we 
have encoded them in the renormalization group functions defined by 
(\ref{gamdef}). Hence, with $\bar{\alpha}$~$=$~$\zeta$~$=$~$0$, we find,  
\begin{eqnarray} 
\gamma_A(a) &=& \frac{1}{6\Noda} \left[ \Noda \left( ( 3 \alpha - 13 ) C_A 
+ 8 T_F \Nf \right) + \Nda \left( ( - 3 \alpha + 9 ) C_A \right) \right] a 
\nonumber \\
&& +~ \frac{1}{48{\Noda}^2} \left[ {\Noda}^2 \left( ( 6 \alpha^2 + 66 \alpha 
- 354 ) C_A^2 + 240 C_A T_F \Nf + 192 C_F T_F \Nf \right) \right. \nonumber \\
&& \left. ~~~~~~~~~~~~~+~ \Noda \Nda \left( ( 3 \alpha^2 + 210 \alpha 
+ 331 ) C_A^2 - 80 C_A T_F \Nf \right) \right. \nonumber \\
&& \left. ~~~~~~~~~~~~~+~ {\Nda}^2 \left( ( 15 \alpha^2 - 6 \alpha - 33 ) 
C_A^2 \right) \right] a^2 \nonumber \\ 
&& +~ \frac{1}{3456{\Noda}^3} \left[ {\Noda}^3 ( ( 162 \alpha^3 + 2727 \alpha^2 
+ 2592 \alpha \zeta_3 + 18036 \alpha + 1944 \zeta_3 - 119580 ) C_A^3 
\right. \nonumber \\
&& \left. ~~~~~~~~~~~~~~~~~~~~~~~ 
+~ ( -~ 6912 \alpha - 62208 \zeta_3 + 174912 ) C_A^2 T_F \Nf 
\right. \nonumber \\
&& \left. ~~~~~~~~~~~~~~~~~~~~~~~ 
+~ ( 82944 \zeta_3 + 960 ) C_A C_F T_F \Nf 
- 29184 C_A T_F^2 \Nf^2 
\right. \nonumber \\
&& \left. ~~~~~~~~~~~~~~~~~~~~~~~ 
-~ 6912 C_F^2 T_F \Nf 
- 16896 ) C_F T_F^2 \Nf^2 ) 
\right. \nonumber \\
&& \left. ~~~~~~~~~~~~~~~~~ 
+~ {\Noda}^2 \Nda ( ( 2133 \alpha^3 + 162 \alpha^2 \zeta_3 + 25785 \alpha^2 
+ 14904 \alpha \zeta_3 + 61479 \alpha 
\right. \nonumber \\
&& \left. ~~~~~~~~~~~~~~~~~~~~~~~~~~~~~~~~~ 
-~ 3564 \zeta_3 + 105550 ) C_A^3 
\right. \nonumber \\
&& \left. ~~~~~~~~~~~~~~~~~~~~~~~~~~~~~~~~~ 
+~ ( -~ 13392 \alpha - 62208 \zeta_3 - 31264 ) C_A^2 T_F \Nf 
\right. \nonumber \\
&& \left. ~~~~~~~~~~~~~~~~~~~~~~~~~~~~~~~~~ 
+~ ( 82944 \zeta_3 - 77760 ) C_A C_F T_F \Nf - 8960 C_A T_F^2 \Nf^2 )
\right. \nonumber \\
&& \left. ~~~~~~~~~~~~~~~~~ 
+~ \Noda {\Nda}^2 ( ( - 324 \alpha^3 \zeta_3 + 1728 \alpha^3 
- 6480 \alpha^2 \zeta_3 + 14256 \alpha^2 
\right. \nonumber \\
&& \left. ~~~~~~~~~~~~~~~~~~~~~~~~~~~~~~~~~ 
- 11988 \alpha \zeta_3 - 26298 \alpha
- 129924 \zeta_3 - 113751 ) C_A^3 
\right. \nonumber \\
&& \left. ~~~~~~~~~~~~~~~~~~~~~~~~~~~~~~~~~ 
+ ( 13392 \alpha + 41472 \zeta_3 + 9936 ) C_A^2 T_F \Nf )
\right. \nonumber \\
&& \left. ~~~~~~~~~~~~~~~~~ 
+ {\Nda}^3 ( ( -~ 4536 \alpha^3 \zeta_3 - 270 \alpha^3 - 18792 \alpha^2 \zeta_3 
- 3294 \alpha^2 - 82296 \alpha \zeta_3 
\right. \nonumber \\
&& \left. ~~~~~~~~~~~~~~~~~~~~~~~~~~~~~~~~~ 
+ 42714 \alpha - 176904 \zeta_3
+ 101952 ) C_A^3 ) \right] a^3 ~+~ O(a^4)
\end{eqnarray}
where $\zeta_n$ is the Riemann zeta function and 
\begin{eqnarray} 
\gamma_\alpha(a) &=& \frac{1}{12 \alpha\Noda} \left[ \Noda \left( 
( -~ 3 \alpha^2 + 26 \alpha ) C_A - 16 \alpha T_F \Nf \right)
+ \Nda \left( ( -~ 6 \alpha^2 - 36 \alpha - 36 ) C_A \right) \right] a 
\nonumber \\ 
&& +~ \frac{1}{48 \alpha {\Noda}^2} \left[ {\Noda}^2 \left( ( -~ 3 \alpha^3 
- 51 \alpha^2 + 354 \alpha ) C_A^2 - 240 \alpha C_A T_F \Nf - 192 \alpha
C_F T_F \Nf \right) \right. \nonumber \\
&& \left. ~~~~~~~~~~~~~~~+~ \Noda \Nda \left( ( -\, 27 \alpha^3 - 339 \alpha^2 
- 647 \alpha - 928 ) C_A^2 + ( 160 \alpha + 512 ) C_A T_F \Nf \right) \right.
\nonumber \\
&& \left. ~~~~~~~~~~~~~~~+~ {\Nda}^2 \left( ( -~ 30 \alpha^3 
- 366 \alpha^2 + 294 \alpha + 2016 ) C_A^2 \right) \right] a^2 \nonumber \\
&& +~ \frac{1}{6912 \alpha {\Noda}^3} \left[ {\Noda}^3 ( ( -~ 162 \alpha^4 
- 3348 \alpha^3 - 5184 \alpha^2 \zeta_3 - 25218 \alpha^2 
\right. \nonumber \\
&& \left. ~~~~~~~~~~~~~~~~~~~~~~~~~~ 
-~ 3888 \alpha \zeta_3 + 239160 \alpha ) C_A^3 
\right. \nonumber \\
&& \left. ~~~~~~~~~~~~~~~~~~~~~~~~~~ 
+~ ( 7344 \alpha^2 + 124416 \alpha \zeta_3 - 349824 \alpha ) C_A^2 T_F \Nf 
\right. \nonumber \\
&& \left. ~~~~~~~~~~~~~~~~~~~~~~~~~~ 
+~ ( -~ 165888 \alpha \zeta_3 - 1920 \alpha ) C_A C_F T_F \Nf 
+ 58368 \alpha C_A T_F^2 \Nf^2 
\right. \nonumber \\
&& \left. ~~~~~~~~~~~~~~~~~~~~~~~~~~ 
+~ 13824 \alpha C_F^2 T_F \Nf + 33792 \alpha C_F T_F^2 \Nf^2 ) 
\right. \nonumber \\
&& \left. ~~~~~~~~~~~~~~~~~~~ 
+~ {\Noda}^2 \Nda ( ( -~ 2754 \alpha^4 - 48492 \alpha^3 
- 14256 \alpha^2 \zeta_3 - 155493 \alpha^2 
\right. \nonumber \\
&& \left. ~~~~~~~~~~~~~~~~~~~~~~~~~~~~~~~~~~~ 
+~ 27864 \alpha \zeta_3 - 256744 \alpha + 209952 \zeta_3 - 548904 ) C_A^3 
\right. \nonumber \\
&& \left. ~~~~~~~~~~~~~~~~~~~~~~~~~~~~~~~~~~~ 
+~ ( 29376 \alpha^2 + 207360 \alpha \zeta_3 + 36064 \alpha 
\right. \nonumber \\
&& \left. ~~~~~~~~~~~~~~~~~~~~~~~~~~~~~~~~~~~~~~~~ 
+~ 331776 \zeta_3 + 136128 ) C_A^2 T_F \Nf 
\right. \nonumber \\
&& \left. ~~~~~~~~~~~~~~~~~~~~~~~~~~~~~~~~~~~ 
+~ ( -~ 331776 \alpha \zeta_3 + 311040 \alpha 
\right. \nonumber \\
&& \left. ~~~~~~~~~~~~~~~~~~~~~~~~~~~~~~~~~~~~~~~~ 
-~ 663552 \zeta_3 + 705024 ) C_A C_F T_F \Nf 
\right. \nonumber \\
&& \left. ~~~~~~~~~~~~~~~~~~~~~~~~~~~~~~~~~~~ 
+~ ( 35840 \alpha + 61440 ) C_A T_F^2 \Nf^2 )
\right. \nonumber \\
&& \left. ~~~~~~~~~~~~~~~~~~~ 
+~ \Noda {\Nda}^2 ( ( -~ 7884 \alpha^4 - 133920 \alpha^3 
+ 76464 \alpha^2 \zeta_3 - 151524 \alpha^2 
\right. \nonumber \\
&& \left. ~~~~~~~~~~~~~~~~~~~~~~~~~~~~~~~~~~~ 
+~ 517752 \alpha \zeta_3 + 503388 \alpha + 1666656 \zeta_3 + 1014012 ) C_A^3 
\right. \nonumber \\
&& \left. ~~~~~~~~~~~~~~~~~~~~~~~~~~~~~~~~~~~ 
+~ ( 29376 \alpha^2 - 248832 \alpha \zeta_3 + 5184 \alpha 
\right. \nonumber \\
&& \left. ~~~~~~~~~~~~~~~~~~~~~~~~~~~~~~~~~~~ 
-~ 995328 \zeta_3 - 812160 ) C_A^2 T_F \Nf  )
\right. \nonumber \\
&& \left. ~~~~~~~~~~~~~~~~~~~ 
+~ {\Nda}^3 ( ( -~ 6480 \alpha^4 - 105840 \alpha^3 - 220320 \alpha^2 \zeta_3 
+ 110700 \alpha^2 
\right. \nonumber \\
&& \left. ~~~~~~~~~~~~~~~~~~~~~~~~~~~~~~~ 
-~ 784080 \alpha \zeta_3 + 373032 \alpha 
\right. \nonumber \\
&& \left. ~~~~~~~~~~~~~~~~~~~~~~~~~~~~~~~ 
-~ 1021248 \zeta_3 - 3148632 ) C_A^3 ) 
\right] a^3 ~+~ O(a^4) ~.  
\end{eqnarray} 
For completeness we record the sum of the previous two anomalous dimensions
partly to indicate the singular nature of this renormalization group function,
but also because it corresponds to the renormalization of the gauge parameter
itself from the convention we have used to define it. We have 
\begin{eqnarray} 
\gamma_A(a) ~+~ \gamma_\alpha(a) &=& \frac{C_A}{4 \alpha\Noda} \left[
\alpha^2 \Noda - \left( 4 \alpha^2 + 6 \alpha + 12 \right) \Nda \right] a 
\nonumber \\ 
&& +~ \frac{C_A}{48 \alpha {\Noda}^2} \left[ {\Noda}^2 \left( ( 3 \alpha^3 
+ 15 \alpha^2 ) C_A \right) \right. \nonumber \\
&& \left. ~~~~~~~~~~~~~~~+~ \Noda \Nda \left( ( -~ 24 \alpha^3 - 129 \alpha^2 
- 316 \alpha - 928 ) C_A \right. \right. \nonumber \\
&& \left. \left. ~~~~~~~~~~~~~~~~~~~~~~~~~~~~~+~ ( 80 \alpha + 512 ) T_F \Nf 
\right) \right. \nonumber \\
&& \left. ~~~~~~~~~~~~~~~+~ {\Nda}^2 \left( ( -~ 15 \alpha^3 
- 372 \alpha^2 + 261 \alpha + 2016 ) C_A \right) \right] a^2 \nonumber \\
&& +~ \frac{1}{6912 \alpha {\Noda}^3} \left[ {\Noda}^3 ( ( 162 \alpha^4 
+ 2106 \alpha^3 + 10854 \alpha^2 ) C_A^3 - 6480 \alpha^2 C_A^2 T_F \Nf )  
\right. \nonumber \\
&& \left. ~~~~~~~~~~~~~~~~~~~ 
+~ {\Noda}^2 \Nda ( ( 1512 \alpha^4 + 324 \alpha^3 \zeta_3 + 3078 \alpha^3 
+ 15552 \alpha^2 \zeta_3 
\right. \nonumber \\
&& \left. ~~~~~~~~~~~~~~~~~~~~~~~~~~~~~~~~~~ 
-~ 32535 \alpha^2 + 20736 \alpha \zeta_3 - 45644 \alpha 
\right. \nonumber \\
&& \left. ~~~~~~~~~~~~~~~~~~~~~~~~~~~~~~~~~~ 
+~ 209952 \zeta_3 - 548904 ) C_A^3 
\right. \nonumber \\
&& \left. ~~~~~~~~~~~~~~~~~~~~~~~~~~~~~~~~~~ 
+~ ( 2592 \alpha^2 + 82944 \alpha \zeta_3 - 26464 \alpha 
\right. \nonumber \\
&& \left. ~~~~~~~~~~~~~~~~~~~~~~~~~~~~~~~~~~~~~~~ 
+~ 331776 \zeta_3 + 136128 ) C_A^2 T_F \Nf 
\right. \nonumber \\
&& \left. ~~~~~~~~~~~~~~~~~~~~~~~~~~~~~~~~~~ 
+~ ( -~ 165888 \alpha \zeta_3 + 155520 \alpha 
\right. \nonumber \\
&& \left. ~~~~~~~~~~~~~~~~~~~~~~~~~~~~~~~~~~~~~~~ 
-~ 663552 \zeta_3 + 705024 ) C_A C_F T_F \Nf 
\right. \nonumber \\
&& \left. ~~~~~~~~~~~~~~~~~~~~~~~~~~~~~~~~~~ 
+~ ( 17920 \alpha + 61440 ) C_A T_F^2 \Nf^2 )
\right. \nonumber \\
&& \left. ~~~~~~~~~~~~~~~~~~~ 
+~ \Noda {\Nda}^2 ( ( -~ 648 \alpha^4 \zeta_3 - 4428 \alpha^4 
- 12960 \alpha^3 \zeta_3 
\right. \nonumber \\
&& \left. ~~~~~~~~~~~~~~~~~~~~~~~~~~~~~~~~~~~
-~ 105408 \alpha^3 + 52488 \alpha^2 \zeta_3 - 204120 \alpha^2 
\right. \nonumber \\
&& \left. ~~~~~~~~~~~~~~~~~~~~~~~~~~~~~~~~~~~
+~ 15552 \alpha^2 \zeta_3 + 257904 \alpha \zeta_3 + 275886 \alpha 
\right. \nonumber \\
&& \left. ~~~~~~~~~~~~~~~~~~~~~~~~~~~~~~~~~~~
+~ 1666656 \zeta_3 + 1014012 ) C_A^3 
\right. \nonumber \\
&& \left. ~~~~~~~~~~~~~~~~~~~~~~~~~~~~~~~~~~~ 
+~ ( 56160 \alpha^2 - 165888 \alpha \zeta_3 + 25056 \alpha 
\right. \nonumber \\
&& \left. ~~~~~~~~~~~~~~~~~~~~~~~~~~~~~~~~~~~~~~~~
-~ 995328 \zeta_3 - 812160 ) C_A^2 T_F \Nf  )
\right. \nonumber \\
&& \left. ~~~~~~~~~~~~~~~~~~~ 
+~ {\Nda}^3 ( ( -~ 9072 \alpha^4 \zeta_3 - 7020 \alpha^4 
- 37584 \alpha^3 \zeta_3 - 112428 \alpha^3 
\right. \nonumber \\
&& \left. ~~~~~~~~~~~~~~~~~~~~~~~~~~~~~~~ 
-~ 384912 \alpha^2 \zeta_3 + 196128 \alpha^2 - 1137888 \alpha \zeta_3 
\right. \nonumber \\
&& \left. ~~~~~~~~~~~~~~~~~~~~~~~~~~~~~~~ 
+~ 576936 \alpha - 1021248 \zeta_3 - 3148632 ) C_A^3 ) 
\right] a^3 \nonumber \\  
&& +~ O(a^4) ~.  
\end{eqnarray} 
Next, 
\begin{eqnarray} 
\gamma_{A^i}(a) &=& \frac{1}{3} \left[ 4 T_F \Nf - 11 C_A \right] a 
\nonumber \\ 
&& +~ \frac{1}{3} \left[ -~ 34 C_A^2 + 20 C_A T_F \Nf + 12 C_F T_F \Nf \right] 
a^2 \nonumber \\
&& +~  \frac{1}{54} \left[ -~ 2857 C_A^3 + 2830 C_A^2 T_F \Nf 
+ 1230 C_A C_F T_F \Nf \right. \nonumber \\
&& \left. ~~~~~~~~-~ 316 C_A T_F^2 \Nf^2 - 108 C_F^2 T_F \Nf 
- 264 C_F T_F^2 \Nf^2 \right] a^3 ~+~ O(a^4) 
\end{eqnarray} 
and we have checked explicitly that when $\bar{\alpha}$ $=$ $0$ 
\begin{equation} 
\gamma_{\alpha^i}(a) ~=~ -~ \gamma_{A^i}(a) ~+~ O(a^4) ~.  
\end{equation} 
For the ghosts and quarks we have 
\begin{eqnarray} 
\gamma_c(a) &=& \frac{1}{4\Noda} \left[ \Noda \left( ( \alpha - 3 ) C_A \right)
+ \Nda \left( ( - 2 \alpha - 6 ) C_A \right) \right] a \nonumber \\
&& +~ \frac{1}{96 {\Noda}^2} \left[ {\Noda}^2 \left( ( 6 \alpha^2 - 6 \alpha 
- 190 ) C_A^2 + 80 C_A T_F \Nf \right) \right. \nonumber \\
&& \left. ~~~~~~~~~~~~~~+~ \Noda \Nda \left( ( -~ 42 \alpha^2 - 126 \alpha 
- 347 ) C_A^2 + 160 C_A T_F \Nf \right) \right. \nonumber \\
&& \left. ~~~~~~~~~~~~~~+~ {\Nda}^2 \left( ( 12 \alpha^2 - 588 \alpha + 510 ) 
C_A^2 \right) \right] a^2 \nonumber \\
&& +~ \frac{1}{6912 {\Noda}^3} \left[ {\Noda}^3 ( ( 162 \alpha^3 
+ 1485 \alpha^2 - 2592 \alpha \zeta_3 + 3672 \alpha - 1944 \zeta_3 
- 63268 ) C_A^3 
\right. \nonumber \\
&& \left. ~~~~~~~~~~~~~~~~~~~~~~~ 
+~ ( -~ 6048 \alpha + 62208 \zeta_3 + 6208 ) C_A^2 T_F \Nf 
\right. \nonumber \\
&& \left. ~~~~~~~~~~~~~~~~~~~~~~~ 
+~ ( -~ 82944 \zeta_3 + 77760 ) C_A C_F T_F \Nf 
+ 8960 C_A T_F^2 \Nf^2 ) 
\right. \nonumber \\
&& \left. ~~~~~~~~~~~~~~~~~ 
+~ {\Noda}^2 \Nda ( ( 1242 \alpha^3 + 10287 \alpha^2 + 8748 \alpha \zeta_3 
+ 2565 \alpha 
\right. \nonumber \\
&& \left. ~~~~~~~~~~~~~~~~~~~~~~~~~~~~~~~~~ 
+~ 57996 \zeta_3 - 19184 ) C_A^3 
\right. \nonumber \\
&& \left. ~~~~~~~~~~~~~~~~~~~~~~~~~~~~~~~~~ 
+~ ( -~ 5616 \alpha + 103680 \zeta_3 - 47632 ) C_A^2 T_F \Nf 
\right. \nonumber \\
&& \left. ~~~~~~~~~~~~~~~~~~~~~~~~~~~~~~~~~ 
+~ ( -~ 165888 \zeta_3 + 155520 ) C_A C_F T_F \Nf 
+ 17920 C_A T_F^2 \Nf^2 ) 
\right. \nonumber \\
&& \left. ~~~~~~~~~~~~~~~~~ 
+~ \Noda {\Nda}^2 ( ( -~ 1296 \alpha^3 \zeta_3 - 1836 \alpha^3 
- 16200 \alpha^2 \zeta_3 - 68148 \alpha^2 
\right. \nonumber \\
&& \left. ~~~~~~~~~~~~~~~~~~~~~~~~~~~~~~~~~ 
+~ 140292 \alpha \zeta_3 - 161730 \alpha + 617868 \zeta_3 - 258174 ) C_A^3 
\right. \nonumber \\
&& \left. ~~~~~~~~~~~~~~~~~~~~~~~~~~~~~~~~~ 
+~ ( 35424 \alpha - 124416 \zeta_3 - 18144 ) C_A^2 T_F \Nf ) 
\right. \nonumber \\
&& \left. ~~~~~~~~~~~~~~~~~ 
+~ {\Nda}^3 ( ( -~ 18144 \alpha^3 \zeta_3 + 864 \alpha^3 
+ 11664 \alpha^2 \zeta_3 - 89532 \alpha^2 - 191160 \alpha \zeta_3 
\right. \nonumber \\
&& \left. ~~~~~~~~~~~~~~~~~~~~~~~~~~~~~ 
+~ 128304 \alpha - 21384 \zeta_3 - 135972 ) C_A^3 ) \right] a^3 ~+~ O(a^4) 
\end{eqnarray} 
\begin{eqnarray} 
\gamma_{c^i}(a) &=& \frac{1}{4 \Noda} \left[ \Noda \left( ( - \alpha - 3 ) C_A
\right) + \Nda \left( ( - 2 \alpha - 6 ) C_A \right) \right] a \nonumber \\
&& +~ \frac{1}{96 {\Noda}^2} \left[ {\Noda}^2 \left( ( -~ 6 \alpha^2 
- 66 \alpha - 190 ) C_A^2 + 80 C_A T_F \Nf \right) \right. \nonumber \\
&& \left. ~~~~~~~~~~~~~~+~ \Noda \Nda \left( ( -~ 54 \alpha^2 - 354 \alpha
- 323 ) C_A^2 + 160 C_A T_F \Nf \right) \right. \nonumber \\
&& \left. ~~~~~~~~~~~~~~+~ {\Nda}^2 \left( ( -~ 60 \alpha^2 - 372 \alpha 
+ 510 ) C_A^2 \right) \right] a^2 \nonumber \\
&& +~ \frac{1}{6912 {\Noda}^3} \left[ {\Noda}^3 ( ( -~ 162 \alpha^3 
- 2727 \alpha^2 - 2592 \zeta_3 \alpha - 18036 \alpha - 1944 \zeta_3 
- 63268 ) C_A^3 
\right. \nonumber \\
&& \left. ~~~~~~~~~~~~~~~~~~~~~~~~ 
+~ ( 6912 \alpha + 62208 \zeta_3 + 6208 ) C_A^2 T_F \Nf 
\right. \nonumber \\
&& \left. ~~~~~~~~~~~~~~~~~~~~~~~~ 
+~ ( -~ 82944 \zeta_3 + 77760 ) C_A C_F T_F \Nf 
+ 8960 C_A T_F^2 \Nf^2 )
\right. \nonumber \\
&& \left. ~~~~~~~~~~~~~~~~~ 
+~ {\Noda}^2 \Nda ( ( -~ 2754 \alpha^3 + 648 \zeta_3 \alpha^2 
- 28917 \alpha^2 - 4212 \zeta_3 \alpha 
\right. \nonumber \\
&& \left. ~~~~~~~~~~~~~~~~~~~~~~~~~~~~~~~~~ 
-~ 69309 \alpha + 37260 \zeta_3 - 64544 ) C_A^3 
\right. \nonumber \\
&& \left. ~~~~~~~~~~~~~~~~~~~~~~~~~~~~~~~~~ 
+~ ( 25488 \alpha + 103680 \zeta_3 - 13072 ) C_A^2 T_F \Nf 
\right. \nonumber \\
&& \left. ~~~~~~~~~~~~~~~~~~~~~~~~~~~~~~~~~ 
+~ ( -~ 165888 \zeta_3 + 155520 ) C_A C_F T_F \Nf 
+ 17920 C_A T_F^2 \Nf^2 )
\right. \nonumber \\
&& \left. ~~~~~~~~~~~~~~~~~ 
+~ \Noda {\Nda}^2 ( ( -~ 7884 \alpha^3 + 22680 \zeta_3 \alpha^2
- 84564 \alpha^2 + 97524 \zeta_3 \alpha
\right. \nonumber \\
&& \left. ~~~~~~~~~~~~~~~~~~~~~~~~~~~~~~~~~ 
-~ 47142 \alpha + 433836 \zeta_3 - 56430 ) C_A^3 
\right. \nonumber \\
&& \left. ~~~~~~~~~~~~~~~~~~~~~~~~~~~~~~~~~ 
+~ ( 25056 \alpha - 124416 \zeta_3 - 18144 ) C_A^2 T_F \Nf )
\right. \nonumber \\
&& \left. ~~~~~~~~~~~~~~~~~ 
+~ {\Nda}^3 ( ( -~ 6480 \alpha^3 + 34992 \zeta_3 \alpha^2 
- 70092 \alpha^2 + 8424 \zeta_3 \alpha 
\right. \nonumber \\
&& \left. ~~~~~~~~~~~~~~~~~~~~~~~~~~~~~ 
+~ 114912 \alpha + 77112 \zeta_3 - 161028 ) C_A^3 ) \right] a^3 ~+~ O(a^4) 
\end{eqnarray} 
and 
\begin{eqnarray} 
\gamma_\psi(a) &=& \frac{\alpha \Noda T_F}{\NF} a \nonumber \\
&& +~ \frac{1}{4\NF} \left[ ( -~ \alpha^2 + 22 \alpha + 23 ) C_A C_F \NF 
+ ( \alpha^2 - 14 \alpha + 2 ) \Noda C_A T_F \right. \nonumber \\
&& \left. ~~~~~~~~~~~-~ 6 C_F^2 \NF - 8 C_F \Nf T_F \NF \right] a^2 
\nonumber \\ 
&& +~ \frac{1}{576 N_F T_F \Noda} \left[ ( 684 \alpha^3 + ( 1296 \zeta_3 
+ 3528 ) \alpha^2 + ( 22464 \zeta_3 - 14094 ) \alpha 
\right. \nonumber \\
&& \left. ~~~~~~~~~~~~~~~~~~~~~~ 
+~ 30240 \zeta_3 - 33264 ) C_A^2 C_F^2 N_F^2 
\right. \nonumber \\
&& \left. ~~~~~~~~~~~~~~~~~~~~~~ 
+~ ( -~ 810 \alpha^3 + ( - 2376 \zeta_3 - 1908 ) \alpha^2
+ ( - 40608 \zeta_3 + 42039 ) \alpha
\right. \nonumber \\
&& \left. ~~~~~~~~~~~~~~~~~~~~~~~~~~~~ 
-~ 63072 \zeta_3 + 109016 ) C_A^2 C_F N_F \Noda T_F 
\right. \nonumber \\
&& \left. ~~~~~~~~~~~~~~~~~~~~~~ 
+~ ( 180 \alpha^3 + ( 1080 \zeta_3 - 1080 ) \alpha^2 
+ ( 18576 \zeta_3 - 23211 ) \alpha 
\right. \nonumber \\
&& \left. ~~~~~~~~~~~~~~~~~~~~~~~~~~~~ 
+~ 27864 \zeta_3 - 39132 ) C_A^2 {\Noda}^2 T_F^2 
\right. \nonumber \\
&& \left. ~~~~~~~~~~~~~~~~~~~~~~ 
+~ ( -~ 5472 \alpha - 24128 ) C_A C_F N_F \Noda \Nf T_F^2 
\right. \nonumber \\
&& \left. ~~~~~~~~~~~~~~~~~~~~~~ 
+~ ( 3024 \alpha + 5760 ) C_A {\Noda}^2 \Nf T_F^3 
\right. \nonumber \\
&& \left. ~~~~~~~~~~~~~~~~~~~~~~ 
+~ ( 6912 \zeta_3 - 20592 ) C_A C_F^2 N_F \Noda T_F 
\right. \nonumber \\
&& \left. ~~~~~~~~~~~~~~~~~~~~~~ 
+~ 1728 C_F^2 N_F \Noda \Nf T_F^2 + 1280 C_F N_F \Noda \Nf^2 T_F^3 
\right. \nonumber \\
&& \left. ~~~~~~~~~~~~~~~~~~~~~~ 
+~ 864 C_F^3 N_F \Noda T_F \right] a^3 ~+~ O(a^4) ~. 
\end{eqnarray} 
Finally, for completeness we record that the $\beta$-function emerges as
\begin{eqnarray}
\beta(a) &=& -~ \left[ \frac{11}{3} C_A - \frac{4}{3} T_F \Nf \right] a^2 ~-~ 
\left[ \frac{34}{3} C_A^2 - 4 C_F T_F \Nf - \frac{20}{3} C_A T_F \Nf \right]
a^3 \nonumber \\  
&& +~ \left[ 2830 C_A^2 T_F \Nf - 2857 C_A^3 + 1230 C_A C_F T_F \Nf 
- 316 C_A T_F^2 \Nf^2 \right. \nonumber \\ 
&& \left. ~~~~~-~ 108 C_F^2 T_F \Nf - 264 C_F T_F^2 \Nf^2 \right] 
\frac{a^4}{54} ~+~ O(a^5) ~.  
\end{eqnarray} 

To have confidence in the correctness of these results it is important to 
indicate the checks we have carried out. First, from a renormalization point of
view, using the method of \cite{5} the double and triple poles in $\epsilon$ at
three loops and the double pole at two loops in the renormalization constants 
are not independent of the previous order one loop poles. Therefore, we have 
checked that these emerge correctly for both the Landau gauge and the MAG. This
is a non-trivial observation given the particular structure of the 
renormalization group functions which depend not only on $\alpha$ but also on 
the colour group Casimirs and for the MAG, the dimensions of the centre and 
off-diagonal sector of the Lie group. Second, the $\beta$-function correctly
emerges from the renormalization of the centre gluon. Again this is non-trivial 
since in the renormalization of the $2$-point function its divergence has to 
emerge to be independent of not only $\alpha$ but also of $\Nda$ and $\Noda$ as
well as being equivalent to the actual $\beta$-function itself. By the same 
token we can of course trivially record that the four loop anomalous dimension 
for $A^i_\mu$ is,
\cite{6},  
\begin{eqnarray} 
\gamma_{A^i}(a) &=& \frac{1}{3} \left[ 4 T_F \Nf - 11 C_A \right] a 
\nonumber \\ 
&& +~ \frac{1}{3} \left[ -~ 34 C_A^2 + 20 C_A T_F \Nf + 12 C_F T_F \Nf \right] 
a^2 \nonumber \\
&& +~  \frac{1}{54} \left[ -~ 2857 C_A^3 + 2830 C_A^2 T_F \Nf 
+ 1230 C_A C_F T_F \Nf \right. \nonumber \\
&& \left. ~~~~~~~~-~ 316 C_A T_F^2 \Nf^2 - 108 C_F^2 T_F \Nf 
- 264 C_F T_F^2 \Nf^2 \right] a^3 \nonumber \\
&& +~ \left[ \left( \frac{44}{9} \zeta_3 - \frac{150653}{486} \right) C_A^4
+ \left( \frac{39143}{81} - \frac{136}{3} \zeta_3 \right) C_A^3 T_F \Nf
\right. \nonumber \\
&& \left. ~~~~~ 
+ \left( \frac{656}{9} \zeta_3 - \frac{7073}{243} \right) C_A^2 C_F T_F \Nf
+ \left( \frac{4204}{27} - \frac{352}{9} \zeta_3 \right) C_A C_F^2 T_F \Nf  
- 46 C_F^3 T_F \Nf 
\right. \nonumber \\
&& \left. ~~~~~ 
- \left( \frac{7930}{81} + \frac{224}{9} \zeta_3 \right) C_A^2 T_F^2 \Nf^2  
+ \left( \frac{704}{9} \zeta_3 - \frac{1352}{27} \right) C_F^2 T_F^2 \Nf^2  
\right. \nonumber \\
&& \left. ~~~~~ 
- \left( \frac{17152}{243} + \frac{448}{9} \zeta_3 \right) C_A C_F T_F^2 \Nf^2 
- \frac{424}{243} C_A T_F^3 \Nf^3 
- \frac{1232}{243} C_F T_F^3 \Nf^3 
\right. \nonumber \\
&& \left. ~~~~~ 
+ \left( \frac{80}{9} - \frac{704}{3} \zeta_3 \right) 
\frac{d_A^{ABCD}d_A^{ABCD}}{\NF}  
+ \left( \frac{1664}{3} \zeta_3 - \frac{512}{9} \right) 
\frac{d_F^{ABCD}d_A^{ABCD}}{\NF} \Nf  
\right. \nonumber \\
&& \left. ~~~~~ 
+ \left( \frac{704}{9} - \frac{512}{3} \zeta_3 \right) 
\frac{d_F^{ABCD}d_F^{ABCD}}{\NF} \Nf^2 \right] a^5  ~+~ O(a^6) 
\end{eqnarray} 
where
\begin{equation}
d_F^{ABCD} ~=~ \mbox{Tr} \left( T^A T^{(B} T^C T^{D)} \right) 
\end{equation}
and $d_A^{ABCD}$ is $d_F^{ABCD}$ evaluated in the adjoint representation.  

The next checks concern the anomalous dimensions themselves in certain limits.
We have already indicated that the programmes we have used correctly reproduce
all the Landau gauge results prior to switching to the MAG. However, the
anomalous dimensions are also related to those of the Curci-Ferrari gauge. For
instance, for the off-diagonal gluon, off-diagonal ghost and quark, taking the
formal limit $\Nda/\Noda$~$\rightarrow$~$0$, then the following anomalous
dimensions arise for {\em arbitrary} $\alpha$, 
\begin{eqnarray} 
\lim_{{\Nda}/{\Noda}\rightarrow 0} 
\gamma_A(a) &=& \frac{1}{6} \left[ 
( 3 \alpha - 13 ) C_A + 8 T_F \Nf \right] a \nonumber \\
&& +~ \frac{1}{48} \left[ ( 6 \alpha^2 + 66 \alpha - 354 ) C_A^2 
+ 240 C_A T_F \Nf + 192 C_F T_F \Nf \right] a^2 \nonumber \\ 
&& +~ \frac{1}{3456} \left[  ( 162 \alpha^3 + 2727 \alpha^2 
+ 2592 \alpha \zeta_3 + 18036 \alpha + 1944 \zeta_3 - 119580 ) C_A^3 
\right. \nonumber \\
&& \left. ~~~~~~~~~~~~ 
+~ ( -~ 6912 \alpha - 62208 \zeta_3 + 174912 ) C_A^2 T_F \Nf \right. 
\nonumber \\
&& \left. ~~~~~~~~~~~~ 
+~ ( 82944 \zeta_3 + 960 ) C_A C_F T_F \Nf - 29184 C_A T_F^2 \Nf^2 
\right. \nonumber \\
&& \left. ~~~~~~~~~~~~ 
-~ 6912 C_F^2 T_F \Nf - 16896 ) C_F T_F^2 \Nf^2 \right] a^3 ~+~ O(a^4)
\end{eqnarray}
\begin{eqnarray} 
\lim_{{\Nda}/{\Noda}\rightarrow 0} 
\gamma_\alpha(a) &=& \frac{1}{12} \left[ ( - 3 \alpha + 26 ) 
C_A - 16 T_F \Nf \right] a \nonumber \\ 
&& +~ \frac{1}{48} \left[ ( -~ 3 \alpha^2 - 51 \alpha + 354 ) 
C_A^2 - 240 C_A T_F \Nf - 192 C_F T_F \Nf \right] a^2 \nonumber \\
&& +~ \frac{1}{6912} \left[ ( -~ 162 \alpha^3 
- 3348 \alpha^2 - 5184 \alpha \zeta_3 - 25218 \alpha 
\right. \nonumber \\
&& \left. ~~~~~~~~~~~~ 
-~ 3888 \zeta_3 + 239160 ) C_A^3 
\right. \nonumber \\
&& \left. ~~~~~~~~~~~~ 
+~ ( 7344 \alpha + 124416 \zeta_3 - 349824 ) C_A^2 T_F \Nf 
\right. \nonumber \\
&& \left. ~~~~~~~~~~~~ 
+~ ( -~ 165888 \zeta_3 - 1920 ) C_A C_F T_F \Nf 
+ 58368 C_A T_F^2 \Nf^2 
\right. \nonumber \\
&& \left. ~~~~~~~~~~~~ 
+~ 13824 C_F^2 T_F \Nf + 33792 C_F T_F^2 \Nf^2 \right] a^3 ~+~ O(a^4)  
\end{eqnarray} 
\begin{eqnarray} 
\lim_{{\Nda}/{\Noda}\rightarrow 0} 
\left[\gamma_A(a) \,+\, \gamma_\alpha(a) \right] &=& \frac{\alpha C_A}{4} a 
+ \frac{\alpha (\alpha+5) C_A^2}{16} a^2 \nonumber \\
&& +~ \frac{3\alpha}{128} \! \left[ \!( \alpha^2 + 13 \alpha + 67 ) C_A 
- 40 T_F \Nf \right] \! C_A^2 a^3 \,+\, O(a^4)  
\end{eqnarray} 
\begin{eqnarray} 
\lim_{{\Nda}/{\Noda}\rightarrow 0} 
\gamma_c(a) &=& \frac{1}{4} \left[ ( \alpha - 3 ) C_A 
\right] a \nonumber \\
&& +~ \frac{1}{96} \left[ ( 6 \alpha^2 - 6 \alpha - 190 ) C_A^2 
+ 80 C_A T_F \Nf \right] a^2 \nonumber \\
&& +~ \frac{1}{6912} \left[ ( 162 \alpha^3 
+ 1485 \alpha^2 - 2592 \alpha \zeta_3 + 3672 \alpha - 1944 \zeta_3 
- 63268 ) C_A^3 
\right. \nonumber \\
&& \left. ~~~~~~~~~~~~ 
+~ ( -~ 6048 \alpha + 62208 \zeta_3 + 6208 ) C_A^2 T_F \Nf 
\right. \nonumber \\
&& \left. ~~~~~~~~~~~~ 
+~ ( -~ 82944 \zeta_3 + 77760 ) C_A C_F T_F \Nf 
+ 8960 C_A T_F^2 \Nf^2 ) \right] a^3 \nonumber \\
&& +~ O(a^4) 
\end{eqnarray} 
and 
\begin{eqnarray} 
\lim_{{\Nda}/{\Noda}\rightarrow 0} 
\gamma_\psi(a) &=& \frac{\alpha C_F}{4} a 
+ \frac{1}{4} \left[ ( 8 \alpha + 25 ) C_A C_F - 6 C_F^2 - 8 C_F T_F \Nf 
\right] a^2 \nonumber \\
&& +~ \frac{1}{288} \left[ ( 27 \alpha^3 + 270 \alpha^2 + 216 \alpha \zeta_3 
+ 2367 \alpha - 2484 \zeta_3 + 18310 ) C_A^2 C_F 
\right. \nonumber \\
&& \left. ~~~~~~~~~~ 
+~ ( -~ 1224 \alpha - 9184 ) C_A C_F T_F \Nf + 432 C_F^3 + 864 C_F T_F \Nf  
\right. \nonumber \\
&& \left. ~~~~~~~~~~ 
+~ ( 3456 \zeta_3 - 10296 ) C_A C_F^2 + 640 C_F T_F^2 \Nf^2 ) \right] a^3 ~+~ 
O(a^4) 
\end{eqnarray} 
where to take the limit for the quark anomalous dimension\footnote{Whilst the
renormalization constant for $Z_\psi$ was explicity given in \cite{12} for the 
Curci-Ferrari gauge, the actual anomalous dimension was inadvertently omitted.}
we have used the result that
\begin{equation}
\lim_{{\Nda}/{\Noda}\rightarrow 0} \frac{T_F \Noda}{\NF} ~=~ C_F ~.  
\end{equation}  
Comparing these limits with \cite{12,51,52}, we observe that they are 
equivalent to the three loop $\MSbar$ anomalous dimensions in the Curci-Ferrari
gauge for arbitrary $\alpha$. That this result appears is not unexpected since
Kondo indicated in \cite{19} that the off-diagonal sector is in fact the
Curci-Ferrari gauge. Indeed this observation, and its relation to the 
generation of a non-zero vacuum expectation value for the operator ${\cal O}$,
was one of the reasons for the recent renewed interest in both the 
Curci-Ferrari gauge and MAG. That the Curci-Ferrari anomalous dimensions
correctly emerge is an important check on the full MAG computation. A final
more trivial check rests in taking the formal abelian limit in the Landau
gauge, $C_A$~$\rightarrow$~$0$, $C_F$~$\rightarrow$~$1$, 
$T_F$~$\rightarrow$~$1$ and $\alpha$~$\rightarrow$~$0$. One observes that both
ghost anomalous dimensions vanish, the centre and off-diagonal gluon anomalous
dimensions reduce to the quantum electrodynamics $\beta$-function and the 
quark anomalous dimension tends to the electron anomalous dimension.

Having justified the results for the full renormalization of the QCD Lagrangian 
in the MAG, we can now deduce the anomalous dimension of ${\cal O}$. At two 
loops we actually computed the anomalous dimension directly by the same method 
as \cite{12}. The operator was inserted in an off-diagonal ghost $2$-point 
function and the corresponding renormalization constant $Z_{\cal O}$ was 
extracted. Computing the associated anomalous dimension directly from the 
renormalization constant, the resulting two loop value correctly satisfied 
(\ref{opid}). At three loops we took the point of view that the three loop 
$\gamma_{c^i}(a)$ was correctly determined and therefore used (\ref{opid}) to 
deduce  
\begin{eqnarray} 
\gamma_{\cal O}(a) &=& \frac{1}{12 \Noda} \left[ \Noda \left( ( -~ 3 \alpha 
+ 35 ) C_A - 16 T_f \Nf \right) + \Nda \left( ( -~ 6 \alpha - 18 ) C_A \right) 
\right] a \nonumber \\
&& +~ \frac{1}{96 {\Noda}^2} \left[ {\Noda}^2 \left(  ( -~ 6 \alpha^2 
- 66 \alpha + 898 ) C_A^2 - 560 C_A T_f \Nf - 384 C_F T_f \Nf \right) \right.
\nonumber \\
&& \left. ~~~~~~~~~~~~~+~ \Noda \Nda \left(  ( -~ 54 \alpha^2 - 354 \alpha 
- 323 ) C_A^2 + 160 C_A T_f \Nf \right) \right. \nonumber \\
&& \left. ~~~~~~~~~~~~~+~ {\Nda}^2 \left(  ( -~ 60 \alpha^2 - 372 \alpha 
+ 510 ) C_A^2 \right) \right] a^2 \nonumber \\
&& +~ \frac{1}{6912 {\Noda}^3} \left[ {\Noda}^3 ( ( -~ 162 \alpha^3 
- 2727 \alpha^2 - 2592 \zeta_3 \alpha - 18036 \alpha - 1944 \zeta_3 
+ 302428 ) C_A^3 
\right. \nonumber \\
&& \left. ~~~~~~~~~~~~~~~~~~~~~~~ 
+~ ( 6912 \alpha + 62208 \zeta_3 - 356032 ) C_A^2 T_F \Nf 
\right. \nonumber \\
&& \left. ~~~~~~~~~~~~~~~~~~~~~~~ 
+~ ( -~ 82944 \zeta_3 - 79680 ) C_A C_F T_F \Nf + 49408 C_A T_F^2 \Nf^2 
\right. \nonumber \\
&& \left. ~~~~~~~~~~~~~~~~~~~~~~~ 
+~ 13824 C_F^2 T_F \Nf + 33792 C_F T_F^2 \Nf^2 )
\right. \nonumber \\
&& \left. ~~~~~~~~~~~~~~~~~ 
+~ {\Noda}^2 {\Nda} ( ( -~ 2754 \alpha^3 + 648 \alpha^2 \zeta_3
- 28917 \alpha^2 - 4212 \alpha \zeta_3 
\right. \nonumber \\
&& \left. ~~~~~~~~~~~~~~~~~~~~~~~~~~~~~~~~~~ 
-~ 69309 \alpha + 37260 \zeta_3 - 64544 ) C_A^3
\right. \nonumber \\
&& \left. ~~~~~~~~~~~~~~~~~~~~~~~~~~~~~~~~~ 
+~ ( 25488 \alpha + 103680 \zeta_3 - 13072 ) C_A^2 T_F \Nf 
\right. \nonumber \\
&& \left. ~~~~~~~~~~~~~~~~~~~~~~~~~~~~~~~~~ 
+~ ( -~ 165888 \zeta_3 + 155520 ) C_A C_F T_F \Nf 
+ 17920 C_A T_F^2 \Nf^2 ) 
\right. \nonumber \\
&& \left. ~~~~~~~~~~~~~~~~~ 
+~ {\Noda} {\Nda}^2 (  ( -~ 7884 \alpha^3 + 22680 \alpha^2 \zeta_3
- 84564 \alpha^2 + 97524 \alpha \zeta_3 - 47142 \alpha 
\right. \nonumber \\
&& \left. ~~~~~~~~~~~~~~~~~~~~~~~~~~~~~~~~~ 
+~ 433836 \zeta_3 - 56430 ) C_A^3
\right. \nonumber \\
&& \left. ~~~~~~~~~~~~~~~~~~~~~~~~~~~~~~~~ 
+~ ( 25056 \alpha - 124416 \zeta_3 - 18144 ) C_A^2 T_F \Nf ) 
\right. \nonumber \\
&& \left. ~~~~~~~~~~~~~~~~~ 
+~ {\Nda}^3 (  ( -~ 6480 \alpha^3 + 34992 \alpha^2 \zeta_3 
- 70092 \alpha^2 + 8424 \alpha \zeta_3 + 114912 \alpha 
\right. \nonumber \\
&& \left. ~~~~~~~~~~~~~~~~~~~~~~~~~~~~~ 
+~ 77112 \zeta_3 - 161028 ) C_A^3) \right] a^3 ~+~ O(a^4) ~.  
\end{eqnarray} 
Again there are several checks on this result aside from the internal
renormalization group consistency check. First, we verified that the Landau
gauge anomalous dimension emerged correctly. Second, in the formal limit
$\Nda/\Noda$~$\rightarrow$~$0$ $\gamma_{\cal O}(a)$ tends to the Curci-Ferrari
gauge result for all $\alpha$, \cite{12,51,52}. 

Finally, having derived the anomalous dimensions of all the fields and 
${\cal O}$ for an arbitrary colour group, we record the explicit results for 
the two main Lie groups of interest. For $SU(2)$ with $C_A$~$=$~$2$, 
$C_F$~$=$~$3/4$, $T_F$~$=$~$1/2$, $\NF$~$=$~$2$, $\Noda$~$=$~$2$ and 
$\Nda$~$=$~$1$, we have  
\begin{eqnarray} 
\gamma_A(a) &=& \left[ 3 \alpha - 17 + 4 \Nf \right] \frac{a}{6} ~+~
\left[ 45 \alpha^2 + 678 \alpha - 787 + 272 \Nf \right] \frac{a^2}{48} 
\nonumber \\ 
&& +~ \left[ -~ 2592 \alpha^3 \zeta_3 + 6507 \alpha^3 - 15552 \alpha^2 \zeta_3 
+ 75087 \alpha^2 - 10260 \alpha \Nf 
\right. \nonumber \\
&& \left. ~~~~
-~ 12960 \alpha \zeta_3 + 190161 \alpha - 10000 \Nf^2 - 36288 \Nf \zeta_3 
+ 146572 \Nf 
\right. \nonumber \\
&& \left. ~~~~
-~ 217728 \zeta_3 - 329995 \right] \frac{a^3}{1728} ~+~ O(a^4) \nonumber \\ 
\gamma_\alpha(a) &=& \left[ -~ 3 \alpha^2 + 4 \alpha - 9 - 2 \alpha \Nf \right]
\frac{a}{3\alpha} \nonumber \\
&& +~ \left[ -~ 12 \alpha^3 - 156 \alpha^2 + 52 \alpha + 20 - ( 29 \alpha - 32 )
\Nf \right] \frac{a^2}{6\alpha} \nonumber \\ 
&& +~ \left[ -~ 2160 \alpha^4 - 37152 \alpha^3 + 3672 \alpha^2 \Nf 
-~ 10368 \alpha^2 \zeta_3 - 63504 \alpha^2 
\right. \nonumber \\
&& \left. ~~~~
+~ 2780 \alpha \Nf^2 + 5184 \alpha \Nf \zeta_3 - 33869 \alpha \Nf 
+ 20736 \alpha \zeta_3 + 141632 \alpha 
\right. \nonumber \\
&& \left. ~~~~
+~ 960 \Nf^2 - 25920 \Nf \zeta_3 - 348 \Nf + 196992 \zeta_3 
- 207264 \right] \frac{a^3}{432 \alpha} ~+~ O(a^4) \nonumber \\ 
\gamma_A(a) ~+~ \gamma_\alpha(a) &=& -~ \left[ \alpha^2 + 3 \alpha + 6 \right]
\frac{a}{2\alpha} \nonumber \\
&& +~ \left[ -~ 51 \alpha^3 - 570 \alpha^2 - 371 \alpha + 160 + ( 40 \alpha 
+ 256 ) \Nf \right] \frac{a^2}{48\alpha} \nonumber \\ 
&& +~ \left[ -~ 2592 \alpha^4 \zeta_3 - 2133 \alpha^4 - 15552 \alpha^3 \zeta_3 
- 73521 \alpha^3 + 4428 \alpha^2 \Nf 
\right. \nonumber \\
&& \left. ~~~~
-~ 54432 \alpha^2 \zeta_3 - 63855 \alpha^2 + 1120 \alpha \Nf^2 
- 15552 \alpha \Nf \zeta_3 + 11096 \alpha \Nf 
\right. \nonumber \\
&& \left. ~~~~
-~ 134784 \alpha \zeta_3 + 236533 \alpha + 3840 \Nf^2 - 103680 \Nf \zeta_3 
- 1392 \Nf 
\right. \nonumber \\
&& \left. ~~~~
+~ 787968 \zeta_3 - 829056 \right] \frac{a^3}{1728 \alpha} ~+~ O(a^4) 
\nonumber \\ 
\gamma_{A^i}(a) &=& 2 \left[ \Nf - 11 \right] \frac{a}{3} ~+~
\left[ 49 \Nf - 272 \right] \frac{a^2}{6} \nonumber \\ 
&& +~ \left[ - 1660 \Nf^2 + 52417 \Nf - 182848 \right] \frac{a^3}{432} ~+~
O(a^4) \nonumber \\ 
\gamma_c(a) &=& -~ 3 \alpha \, a ~+~ \left[ -~ 3 \alpha^2 - 54 \alpha - 59 
+ 10 \Nf \right] \frac{a^2}{6} \nonumber \\ 
&& +~ \left[ -~ 648 \alpha^3 \zeta_3 + 108 \alpha^3 - 648 \alpha^2 \zeta_3 
- 5400 \alpha^2 + 3240 \alpha \zeta_3 - 4860 \alpha + 280 \Nf^2 
\right. \nonumber \\
&& \left. ~~~~
+~ 1296 \Nf \zeta_3 + 2261 \Nf + 44712 \zeta_3 - 38600 \right] 
\frac{a^3}{216} ~+~ O(a^4) \nonumber \\ 
\gamma_{c^i}(a) &=& -~ \left[ \alpha + 3 \right] a ~+~ \left[ -~ 6 \alpha^2 
- 42 \alpha - 28 + 5 \Nf \right] \frac{a^2}{3} \nonumber \\ 
&& +~ \left[ -~ 1080 \alpha^3 + 2592 \alpha^2 \zeta_3 - 11772 \alpha^2 
+ 1620 \alpha \Nf + 5184 \alpha \zeta_3 - 12528 \alpha  
\right. \nonumber \\
&& \left. ~~~~
+~ 280 \Nf^2 + 1296 \Nf \zeta_3 + 3341 \Nf + 33696 \zeta_3 - 32444 \right]
 \frac{a^3}{216} ~+~ O(a^4) \nonumber \\ 
\gamma_\psi(a) &=& \frac{1}{2} \alpha \, a ~+~ \left[ -~ 4 \alpha^2 + 152 
\alpha + 265 - 24 \Nf \right] \frac{a^2}{32} ~+~ O(a^3) \nonumber \\ 
&& +~ \left[ 672 \alpha^3 + 576 \alpha^2 \zeta_3 + 5328 \alpha^2 
- 1728 \alpha \Nf + 10944 \alpha \zeta_3 + 10848 \alpha 
\right. \nonumber \\
&& \left. ~~~~
+~ 160 \Nf^2 - 9820 \Nf + 6912 \zeta_3 + 50863 \right] 
\frac{a^3}{384} ~+~ O(a^4)
\nonumber \\ 
\gamma_{\cal O}(a) &=& \left[ -~ 3 \alpha + 13 - 2 \Nf \right] \frac{a}{3} ~+~
\left[ -~ 4 \alpha^2 - 28 \alpha + 72 - 13 \Nf \right] \frac{a^2}{2}  
\nonumber \\ 
&& +~ \left[ -~ 720 \alpha^3 + 1728 \alpha^2 \zeta_3 - 7848 \alpha^2 
+ 1080 \alpha \Nf + 3456 \alpha \zeta_3 - 8352 \alpha 
\right. \nonumber \\
&& \left. ~~~~
+~ 740 \Nf^2 + 864 \Nf \zeta_3 - 15245 \Nf + 22464 \zeta_3 + 39320 \right] 
\frac{a^3}{144} ~+~ O(a^4) ~. \nonumber \\  
\end{eqnarray} 
The one loop expressions agree with the limited known results. Repeating the 
same exercise for $SU(3)$ with $C_A$~$=$~$3$, $C_F$~$=$~$4/3$, $T_F$~$=$~$1/2$,
$\NF$~$=$~$3$, $\Noda$~$=$~$6$ and $\Nda$~$=$~$2$, we have  
\begin{eqnarray} 
\gamma_A(a) &=& \left[ 3 \alpha - 15 + 2 \Nf \right] \frac{a}{3} ~+~
\left[ 39 \alpha^2 + 609 \alpha - 1113 + 224 \Nf \right] \frac{a^2}{24} 
\nonumber \\ 
&& +~ \left[ -~ 1836 \alpha^3 \zeta_3 + 9495 \alpha^3 - 12258 \alpha^2 \zeta_3 
+ 115056 \alpha^2 - 14832 \alpha \Nf 
\right. \nonumber \\
&& \left. ~~~~
+~ 28620 \alpha \zeta_3 + 334701 \alpha - 9920 \Nf^2 - 43776 \Nf \zeta_3 
+ 229704 \Nf 
\right. \nonumber \\
&& \left. ~~~~
-~ 182088 \zeta_3 - 839337 \right] \frac{a^3}{1152} ~+~ O(a^4) \nonumber \\ 
\gamma_\alpha(a) &=& \left[ -~ 15 \alpha^2 + 42 \alpha - 36 - 8 \alpha \Nf 
\right] \frac{a}{12\alpha} \nonumber \\
&& +~ \left[ -~ 138 \alpha^3 - 1842 \alpha^2 + 1539 \alpha - 768 
- ( 408 \alpha - 256 ) \Nf \right] \frac{a^2}{48\alpha} \nonumber \\ 
&& +~ \left[ -~ 59292 \alpha^4 - 1034424 \alpha^3 + 91800 \alpha^2 \Nf 
- 259200 \alpha^2 \zeta_3 - 2424195 \alpha^2 
\right. \nonumber \\
&& \left. ~~~~
+~ 64000 \alpha \Nf^2 + 193536 \alpha \Nf \zeta_3 - 1301712 \alpha \Nf 
+ 914976 \alpha \zeta_3 
\right. \nonumber \\
&& \left. ~~~~
+~ 6029820 \alpha + 15360 \Nf^2 - 442368 \Nf \zeta_3 + 268128 \Nf 
\right. \nonumber \\
&& \left. ~~~~
+~ 5868288 \zeta_3 - 5046732 \right] \frac{a^3}{6912 \alpha} ~+~ O(a^4)
\nonumber \\ 
\gamma_A(a) ~+~ \gamma_\alpha(a) &=& -~ \left[ \alpha^2 + 6 \alpha + 12 \right]
\frac{a}{4\alpha} \nonumber \\
&& +~ \left[ -~ 60 \alpha^3 - 624 \alpha^2 - 687 \alpha - 768 + ( 40 \alpha 
+ 256 ) \Nf \right] \frac{a^2}{48\alpha} ~+~ O(a^3) \nonumber \\ 
&& +~ \left[ -~ 11016 \alpha^4 \zeta_3 - 2322 \alpha^4 - 73548 \alpha^3 \zeta_3 
- 344088 \alpha^3 + 2808 \alpha^2 \Nf 
\right. \nonumber \\
&& \left. ~~~~
-~ 87480 \alpha^2 \zeta_3 - 415989 \alpha^2 + 4480 \alpha \Nf^2 
- 69120 \alpha \Nf \zeta_3 + 76512 \alpha \Nf 
\right. \nonumber \\
&& \left. ~~~~
-~ 177552 \alpha \zeta_3 + 993798 \alpha + 15360 \Nf^2 - 442368 \Nf \zeta_3 
+ 268128 \Nf 
\right. \nonumber \\
&& \left. ~~~~
+~ 5868288 \zeta_3 - 5046732 \right] \frac{a^3}{6912 \alpha} ~+~ O(a^4) 
\nonumber \\ 
\gamma_{A^i}(a) &=& \left[ 2 \Nf - 33 \right] \frac{a}{3} ~+~
2 \left[ 19 \Nf - 153 \right] \frac{a^2}{3} \nonumber \\ 
&& +~ \left[ -~ 325 \Nf^2 + 15099 \Nf - 77139 \right] \frac{a^3}{54} ~+~ O(a^4)
\nonumber \\ 
\gamma_c(a) &=& \left[ \alpha - 15 \right] \frac{a}{4} ~+~ 
\left[ -~ 60 \alpha^2 - 1020 \alpha - 2241 + 200 \Nf \right] \frac{a^2}{96} 
\nonumber \\ 
&& +~ \left[ -~ 22032 \alpha^3 \zeta_3 + 10908 \alpha^3 
- 36936 \alpha^2 \zeta_3 - 161298 \alpha^2 - 17928 \alpha \Nf 
\right. \nonumber \\
&& \left. ~~~~
+~ 238464 \alpha \zeta_3 - 234657 \alpha + 11200 \Nf^2 + 96768 \Nf \zeta_3 
+ 206616 \Nf 
\right. \nonumber \\
&& \left. ~~~~
+~ 2301696 \zeta_3 - 2791386 \right] \frac{a^3}{6912} ~+~ O(a^4)
\nonumber \\ 
\gamma_{c^i}(a) &=& -~ 5 \left[ \alpha + 3 \right] \frac{a}{4} ~+~ 
\left[ -~ 276 \alpha^2 - 2028 \alpha - 2169 + 200 \Nf \right] \frac{a^2}{96} 
\nonumber \\ 
&& +~ \left[ -~ 59292 \alpha^3 + 108864 \alpha^2 \zeta_3 - 657666 \alpha^2 
+ 81864 \alpha \Nf + 193104 \alpha \zeta_3 
\right. \nonumber \\
&& \left. ~~~~
-~ 1137267 \alpha + 11200 \Nf^2 + 96768 \Nf \zeta_3 + 258456 \Nf 
\right. \nonumber \\
&& \left. ~~~~
+~ 1661472 \zeta_3 - 2619450 \right] \frac{a^3}{6912} ~+~ O(a^4) \nonumber \\  
\gamma_\psi(a) &=& \alpha \, a ~+~ \left[ -~ 3 \alpha^2 + 138 \alpha + 262 
- 16 \Nf \right] \frac{a^2}{12} ~+~ O(a^3) \nonumber \\ 
&& +~ \left[ 8532 \alpha^3 + 5832 \alpha^2 \zeta_3 + 71496 \alpha^2 
- 19224 \alpha \Nf + 117936 \alpha \zeta_3 + 210195 \alpha 
\right. \nonumber \\
&& \left. ~~~~
+~ 1280 \Nf^2 - 114240 \Nf + 43848 \zeta_3 + 948012 \right] 
\frac{a^3}{1728} ~+~ O(a^4) 
\nonumber \\ 
\gamma_{\cal O}(a) &=& \left[ -~ 15 \alpha + 87 - 8 \Nf \right] \frac{a}{12}
\nonumber \\
&& +~ \left[ -~ 276 \alpha^2 - 2028 \alpha + 7623 - 1016 \Nf \right] 
\frac{a^2}{96} \nonumber \\ 
&& +~ \left[ -~ 19764 \alpha^3 + 36288 \alpha^2 \zeta_3 - 219222 \alpha^2 
+ 27288 \alpha \Nf + 64368 \alpha \zeta_3 
\right. \nonumber \\
&& \left. ~~~~
-~ 379089 \alpha + 17600 \Nf^2 + 32256 \Nf \zeta_3 - 558072 \Nf 
\right. \nonumber \\
&& \left. ~~~~
+~ 553824 \zeta_3 + 2418114 \right] \frac{a^3}{2304} ~+~ O(a^4) ~.  
\end{eqnarray} 

\sect{Discussion.} 
We have provided a comprehensive discussion on the three loop $\MSbar$ 
renormalization of QCD in the maximal abelian gauge. Indeed this article
represents the first calculations beyond {\em one} loop as well as the first 
for Lie groups other than just $SU(2)$. By explicit computation we have 
determined all the anomalous dimensions and $\beta$-function before deducing 
the anomalous dimension of ${\cal O}$ at three loops in $\MSbar$. Indeed it is 
the explicit expression for the latter which will be the key to studies of the 
condensation of ${\cal O}$ in the MAG which we hope to examine next. One useful
observation from the main results is the relation of the MAG anomalous 
dimensions to those of other gauges and in particular the Curci-Ferrari gauge. 
That the results for the latter appear in the formal limit 
$\Nda/\Noda$~$\rightarrow$~$0$ is reassuring, though their prior existence was 
also of a more practical use in helping to establish the veracity of the final 
MAG renormalization group functions. Though from the actual structure of the 
final expressions it is clear that they could not be constructed from knowledge
of the same anomalous dimensions in the Landau or Curci-Ferrari gauges. 

\vspace{1cm}
\noindent
{\bf Acknowledgements.} The author thanks Prof S. Sorella, D. Dudal and R.E.
Browne for useful discussions. The calculations were performed with the help of
the computer algebra and symbolic manipulation programme {\sc Form}, \cite{8}.

\appendix

\sect{Feynman rules.} 
In this appendix we record the Feynman rules we used for the maximal abelian
gauge fixing in momentum space which are derived from (\ref{maglag1}) and 
(\ref{maglag}) using a symbolic manipulation programme written in {\sc Form}. 
For the propagators we have 
\begin{eqnarray} 
\langle A^a_\mu(p) A^b_\nu(-p) \rangle 
&=& -~ \frac{\delta^{ab}}{p^2} \left[ \eta_{\mu\nu} ~-~ (1-\alpha) 
\frac{p_\mu p_\nu}{p^2} \right] \nonumber \\ 
\langle A^i_\mu(p) A^j_\nu(-p) \rangle 
&=& -~ \frac{\delta^{ij}}{p^2} \left[ \eta_{\mu\nu} ~-~ (1-\bar{\alpha}) 
\frac{p_\mu p_\nu}{p^2} \right] \nonumber \\ 
\langle c^a(p) {\bar c}^b(-p) \rangle 
&=& \frac{\delta^{ab}}{p^2} \nonumber \\  
\langle c^i(p) {\bar c}^j(-p) \rangle 
&=& \frac{\delta^{ij}}{p^2} \nonumber \\  
\langle \psi(p) {\bar \psi}(-p) \rangle 
&=& \frac{\pslash}{p^2} 
\end{eqnarray} 
where $p$ is the momentum. The non-zero $3$- and $4$-point vertices are 
\begin{eqnarray} 
\langle A^a_\mu(p_1) \bar{\psi}(p_2) \psi(p_3) \rangle 
&=& g T^a \gamma_\mu  
\nonumber \\ 
\langle A^i_\mu(p_1) \bar{\psi}(p_2) \psi(p_3) \rangle 
&=& g T^i \gamma_\mu  
\nonumber \\ 
\langle A^a_\mu(p_1) \bar{c}^b(p_2) c^c(p_3) \rangle 
&=& -~ i g f^{abc} \left( - \frac{1}{2} p_1 - p_3 \right)_\mu 
\nonumber \\ 
\langle A^a_\mu(p_1) \bar{c}^b(p_2) c^k(p_3) \rangle 
&=& -~ i g f^{abk} \left( - \zeta p_3 \right)_\mu  
\nonumber \\ 
\langle A^a_\mu(p_1) \bar{c}^j(p_2) c^c(p_3) \rangle 
&=& -~ i g f^{acj} \left( p_1 + p_3 \right)_\mu  
\nonumber \\ 
\langle A^i_\mu(p_1) \bar{c}^b(p_2) c^c(p_3) \rangle 
&=& -~ i g f^{bci} \left( - p_1 - 2 p_3 + p_3 \zeta \right)_\mu  
\nonumber \\ 
\langle A^a_\mu(p_1) A^b_\nu(p_2) A^c_\sigma(p_3) \rangle 
&=& i g f^{abc} \left( \eta_{\nu\sigma} (p_2 - p_3)_\mu 
                       + \eta_{\sigma\mu} (p_3 - p_1)_\nu 
                       + \eta_{\mu\nu} (p_1 - p_2)_\sigma \right)   
\nonumber \\ 
\langle A^a_\mu(p_1) A^b_\nu(p_2) A^c_\sigma(p_3) A^d_\rho(p_4) \rangle 
&=& -~ \left[ f_d^{abcd}  \left(
          - \eta_{\mu\sigma} \eta_{\nu\rho}
          + \eta_{\mu\rho} \eta_{\nu\sigma}
          \right)
       + f_d^{acbd}  \left(
          - \eta_{\mu\nu} \eta_{\sigma\rho}
          + \eta_{\mu\rho} \eta_{\nu\sigma}
          \right)
\right. \nonumber \\
&& \left. ~~~~~ 
       + f_d^{adbc}  \left(
          - \eta_{\mu\nu} \eta_{\sigma\rho}
          + \eta_{\mu\sigma} \eta_{\nu\rho}
          \right)
       + f_o^{abcd}  \left(
          - \eta_{\mu\sigma} \eta_{\nu\rho}
          + \eta_{\mu\rho} \eta_{\nu\sigma}
          \right)
\right. \nonumber \\
&& \left. ~~~~~ 
       + f_o^{acbd}  \left(
          - \eta_{\mu\nu} \eta_{\sigma\rho}
          + \eta_{\mu\rho} \eta_{\nu\sigma}
          \right)
       + f_o^{adbc}  \left(
          - \eta_{\mu\nu} \eta_{\sigma\rho}
          + \eta_{\mu\sigma} \eta_{\nu\rho}
          \right) 
    \right] 
\nonumber \\ 
\langle A^a_\mu(p_1) A^b_\nu(p_2) A^c_\sigma(p_3) A^l_\rho(p_4) \rangle 
&=& -~ g \left( f_o^{abcl}  \left(
         - \eta_{\mu\sigma} \eta_{\nu\rho}
         + \eta_{\mu\rho} \eta_{\nu\sigma}
         \right)
\right. \nonumber \\
&& \left. ~~~~~~ 
      + f_o^{acbl}  \left(
         - \eta_{\mu\nu} \eta_{\sigma\rho}
         + \eta_{\mu\rho} \eta_{\nu\sigma}
         \right)
\right. \nonumber \\
&& \left. ~~~~~~ 
      + f_o^{albc}  \left(
         - \eta_{\mu\nu} \eta_{\sigma\rho}
         + \eta_{\mu\sigma} \eta_{\nu\rho}
         \right) 
    \right) 
\nonumber \\ 
\langle A^a_\mu(p_1) A^b_\nu(p_2) A^k_\sigma(p_3) A^l_\rho(p_4) \rangle 
&=& -~ g \left(  f_o^{akbl}  \left(
         - \eta_{\mu\nu} \eta_{\sigma\rho}
         + \frac{\zeta(2-\zeta)}{2\alpha} \eta_{\mu\sigma} \eta_{\nu\rho} 
         - \frac{1}{2\alpha} \eta_{\mu\sigma} \eta_{\nu\rho} 
\right. \right. \nonumber \\
&& \left. \left. ~~~~~~~~~~~~~~~ 
         + \eta_{\mu\rho} \eta_{\nu\sigma}
         \right)
\right. \nonumber \\
&& \left. ~~~~~~ 
      + f_o^{albk}  \left(
         - \eta_{\mu\nu} \eta_{\sigma\rho}
         + \eta_{\mu\sigma} \eta_{\nu\rho}
         + \frac{\zeta(2-\zeta)}{2\alpha} \eta_{\mu\rho} \eta_{\nu\sigma}
\right. \right. \nonumber \\
&& \left. \left. ~~~~~~~~~~~~~~~~~ 
         - \frac{1}{2\alpha} \eta_{\mu\rho} \eta_{\nu\sigma} 
         \right)
\right. \nonumber \\
&& \left. ~~~~~~ 
      + f_o^{bkal}  \left(
          \frac{\zeta(2-\zeta)}{2\alpha} \eta_{\mu\rho} \eta_{\nu\sigma} 
         - \frac{1}{2\alpha} \eta_{\mu\rho} \eta_{\nu\sigma} 
         \right)
\right. \nonumber \\
&& \left. ~~~~~~ 
      + f_o^{blak}  \left(
          \frac{\zeta(2-\zeta)}{2\alpha} \eta_{\mu\sigma} \eta_{\nu\rho} 
         - \frac{1}{2\alpha} \eta_{\mu\sigma} \eta_{\nu\rho} 
         \right)
    \right) 
\nonumber \\ 
\langle A^a_\mu(p_1) A^b_\nu(p_2) \bar{c}^c(p_3) c^d(p_4) \rangle 
&=& -~ g \left(  f_d^{acbd}  \left(
         - \eta_{\mu\nu}
         + \zeta \eta_{\mu\nu} 
         \right)
      + f_d^{bcad}  \left(
         - \eta_{\mu\nu}
         + \zeta\eta_{\mu\nu} 
         \right)
    \right) 
\nonumber \\ 
\langle A^a_\mu(p_1) A^j_\nu(p_2) \bar{c}^c(p_3) c^d(p_4) \rangle 
&=& -~ g \left( f_o^{adcj}  \left(
         - \eta_{\mu\nu}
         + \zeta\eta_{\mu\nu} 
         \right)
      + f_o^{ajcd}  \left(
          \frac{1}{2} \eta_{\mu\nu}
         - \frac{\zeta}{2} \eta_{\mu\nu} \zeta
         \right)
    \right) 
\nonumber \\ 
\langle A^a_\mu(p_1) A^j_\nu(p_2) \bar{c}^c(p_3) c^l(p_4) \rangle 
&=& -~ g \left( f_o^{ajcl}  \left(
          \eta_{\mu\nu}
         - \zeta\eta_{\mu\nu} 
         \right)
      + f_o^{alcj}  \left(
         - \zeta\eta_{\mu\nu}
         + \eta_{\mu\nu} 
         \right)
    \right) 
\nonumber \\ 
\langle A^i_\mu(p_1) A^j_\nu(p_2) \bar{c}^c(p_3) c^d(p_4) \rangle 
&=& -~ g \left( f_o^{cidj}  \left(
          \eta_{\mu\nu}
         - \zeta\eta_{\mu\nu} 
         \right)
      + f_o^{cjdi}  \left(
          \zeta\eta_{\mu\nu}
         - \eta_{\mu\nu}
         \right)
    \right) 
\nonumber \\ 
\langle \bar{c}^a(p_1) c^b(p_2) \bar{c}^c(p_3) c^d(p_4) \rangle 
&=& -~ g \left( \alpha f_d^{acbd}  
      - \frac{\alpha}{4} f_o^{abcd} 
      + \frac{\alpha}{2} f_o^{acbd} 
      - \frac{\alpha}{4} f_o^{adbc} 
    \right) 
\nonumber \\ 
\langle \bar{c}^a(p_1) c^b(p_2) \bar{c}^c(p_3) c^l(p_4) \rangle 
&=& -~ g \left( - \frac{\alpha}{2} f_o^{abcl} 
      + \frac{\alpha}{2} f_o^{acbl} 
      - \frac{\alpha}{2} f_o^{albc} 
    \right) 
\nonumber \\ 
\langle \bar{c}^a(p_1) c^j(p_2) \bar{c}^c(p_3) c^l(p_4) \rangle 
&=& -~ g \left( - \alpha f_o^{ajcl} 
      + \alpha f_o^{alcj} 
    \right) 
\label{feynrule} 
\end{eqnarray} 
where the momentum flow for each field is into the vertex. We note that we have
recorded the Feynman rules as generated from the full MAG Lagrangian, using a
{\sc Form} routine, without recourse to the simplifying properties of the 
Jacobi identity of the Lie algebra. For example, the final rule of the
non-zero set, (\ref{feynrule}), actually vanishes after application of a Jacobi
identity. However, in the construction of the routines to perform the overall 
calculation, we have relegated all the algebra associated with the group theory
to a common {\sc Form} module which encodes the necessary simplifying lemmas.  
That module is placed after the module where the above Feynman rules are 
substituted. The remaining Feynman rules for the $3$- and $4$-point vertices 
which have not been recorded above are trivially zero due to the fact that they
would involve either two or more centre indices in the case of the $3$-point
vertices or three or more centre indices for the $4$-point vertices.

\end{document}